\documentclass[11pt,a4paper]{article}
\usepackage[]{fontenc}

\usepackage{latexsym}
\usepackage{amsmath,amsthm,amsfonts,amssymb,amscd,ulem}
\usepackage[dvips]{graphicx}

\usepackage{enumerate}

 \theoremstyle{plain}    
 \newtheorem{thm}{Theorem}[section]
 \numberwithin{equation}{section} 
 \numberwithin{figure}{section} 
 \theoremstyle{plain}
 \theoremstyle{definition}
 \newtheorem{defn}[thm]{Definition}
 \theoremstyle{plain}    
 \newtheorem{lem}[thm]{Lemma} 
 \theoremstyle{plain}    
 \newtheorem{cor}[thm]{Corollary} 
 \theoremstyle{plain}    
 \newtheorem{prop}[thm]{Proposition} 
 \theoremstyle{remark}
 \newtheorem{rem}[thm]{Remark}

\makeatletter
 
 \@addtoreset{equation}{section}
\makeatother

\renewcommand{\qed}{ \hfill \hbox{\rule[-2pt]{3pt}{6pt}}}

\def\rnum#1{\expandafter{\romannumeral #1}}
\def\Rnum#1{\uppercase\expandafter{\romannum}}

\makeatother
\begin{document}
\thispagestyle{empty}

\ 

\vspace{15ex}
\begin{center}\textbf{\huge On implementability of Bogoliubov automorphisms
and the white noise distribution theory for the Fermion system}\end{center}{\huge \par}
\vspace{20ex}

\begin{center}{\large Yoshihito Shimada}\end{center}{\large \par}

\begin{center}\textit{Graduate School of Mathematics}\end{center}

\begin{center}\textit{Kyushu University}\end{center}

\begin{center}\textit{1-10-6 Hakozaki, Fukuoka 812-8581}\end{center}

\begin{center}\textit{JAPAN}\end{center}

\vspace{20ex}
$\,$

In this paper, we consider Bogoliubov automorphisms of CAR algebra
and we give implementers as continuous linear operators on the white
noise (test or generalized) functionals for the Fermion system with
the help of the Fock expansion. 

\vspace{1ex}
\hrule
\vspace{1ex}

\textbf{KEY WORDS:} white noise calculus, Fermion system, Fock expansion,
Bogoliubov automorphism, implementer

\texttt{\textbf{\small e-mail: shimada@math.kyushu-u.ac.jp}}{\small \par}

\newpage

\section{Introduction}

The purpose of this paper is to show the usefulness of white noise
distribution theory for the Fermion system via implementability of
Bogoliubov automorphisms.

First, we explain the Bogoliubov automorphisms. Let $K$ be a complex
vector space with an inner product or a symplectic form. Let $U$
be a bijection on $K$ preserving the inner product or the symplectic
form, and commuting with the anti-linear involution on $K$. We call
$U$ a Bogoliubov transformation. Bogoliubov transformations yield
automorphisms on CAR algebra or CCR algebra. (Here CAR stands for
{}``canonical anti-commutation relations'', and CCR stands for {}``canonical
commutation relations''. ) These automorphisms are called Bogoliubov
automorphisms. 

There are many studies on implementability of Bogoliubov automorphisms.
For example, H. Araki \cite{Araki_Contemp_math} (resp. Matsui and
Shimada \cite{Matsui+Shimada}) is an study of unitary implementers
of Bogoliubov automorphisms for CAR algebra (resp. CCR algebra). The
criterion for implementability of Bogoliubov automorphisms is well-known
as the Hilbert-Schmidt condition. (See Theorem \ref{thm:implementability_condition}
(iii) of this article.) A. Carey and S. Ruijsenaars \cite{Carey+Ruisenaars}
is another example of the study on unitary implementer of Bogoliubov
automorphisms. They deal with Bogoliubov automorphisms constructed
from the loop group.

However, it is not enough for us to look for implementers in unitary,
or bounded operators. Ruijsenaars \cite{Ruijsenaars} deals with Bogoliubov
transformations determined by maps from $\mathbf{R}^{2n-1}$($n>1$)
to unitary matrices. In \cite{Ruijsenaars}, Ruijsenaars shows that
these automorphisms are not unitary implementable except for the trivial
case, that is, almost all Bogoliubov automorphisms violate the Hilbert-Schmidt
conditions. Therefore Ruijsenaars gives implementers as quadratic
forms in \cite{Ruijsenaars}. 

Another studies of {}``extended'' implementers are A. Carey and
J. Palmer \cite{Carey+Palmer} and P. Kristensen \cite{Kristensen III}.
Carey and Palmer \cite{Carey+Palmer} gives implementers as unbounded
linear operators and Kristensen \cite{Kristensen III} realizes implementers
as linear operators on distributions. 

In this paper, we consider Bogoliubov automorphisms determined by
one particle representation of gauge group $C^{\infty}(\mathbf{T}^{2n-1},SO(2))$,
$n\geq1$, and we give implementers in terms of the white noise distribution
theory for the Fermion system.

Next, we describe the white noise calculus for the Fermion system.
Our white noise calculus is the theory for (test or generalized) functionals
on the infinite dimensional space, for continuous linear operators
on these functionals. The white noise distribution theory for the
Boson system was introduced by T. Hida in 1975 and is applied to various
fields, for example, stochastic differential equations, harmonic analysis
on the infinite dimensional space, (infinite dimensional) group representation
theory, topology, mathematical physics, and so forth. In particular,
from the viewpoint of the operator theory, the white noise distribution
theory has the powerful tool called the Fock expansion. The Fock expansion
is the series of integral kernel operators for any continuous linear
operators on the white noise (test or generalized) functionals. For
example, the Fock expansion is applied to determining the rotation
invariant operators on the white noise test functionals \cite{Obata-paper-rotation_inv},
and proving irreducibility of energy representation of gauge group
\cite{Shimada}. 

The white noise calculus for the Fermion system is discussed in \cite{Shimada 2}
and the author shows the Fock expansion for the Fermion system, and
as we said at the beginning of this section, we show usefulness of
the Fock expansion for the Fermion system through the discussion for
implementability of Bogoliubov automorphisms.

This paper is organized as follows. In section 2, we review the white
noise distribution theory for the Fermion system. In section 3, we
define Bogoliubov automorphisms and we give the criterion for implementability
of Bogoliubov automorphisms. This criterion is the main theorem of
this paper. In section 4, we prove the main theorem of this paper
and we obtain implementer by direct computation with the help of the
Fock expansion for the Fermion system. In section 5, we give examples
of Bogoliubov automorphisms and see implementability of these Bogoliubov
automorphisms. Here we deal with Bogoliubov automorphisms determined
by one particle representation of gauge algebra $C^{\infty}(\mathbf{T}^{2n-1},so(2))$,
$n\geq1$.

\section{White noise distribution theory}

In this section, we review the white noise calculus for the Fermion
system. 

\begin{defn}
\label{thm:property_of_self-adj_op_A}Let $H$ be a complex Hilbert
space with an inner product $(\cdot,\cdot)_{0}$. Let $A$ be a self-adjoint
operator defined on a dense domain $D(A)$. Let $\{\lambda_{j}\}_{j\in\mathbf{N}}$
be eigenvalues of $A$ and $\{ e_{j}\}_{j\in\mathbf{N}}$ be normalized
eigenvectors for $\{\lambda_{j}\}_{j\in\mathbf{N}}$, i.e., $Ae_{j}=\lambda_{j}e_{j}$
, $j\in\mathbf{N}$. Moreover, we also assume the following two conditions
:

(i) $\{ e_{j}\}_{j\in\mathbf{N}}$ is a C.O.N.S. of $H$, 

(ii) Multiplicity of $\{\lambda_{j}\}_{j\in\mathbf{N}}$ is finite
and $1<\lambda_{1}\leq\lambda_{2}\leq\ldots\rightarrow\infty$.

{\noindent}Then we have the following properties.
\begin{enumerate}
\item For $p\in\mathbf{R}_{\geq0}$ and $x,\, y\in D(A^{p})$, let $(x,y)_{p}:=(A^{p}x,A^{p}y)_{0}$.
Then $(\cdot,\cdot)_{p}$ is an inner product on $D(A^{p})$. Moreover,
$D(A^{p})$ is complete with respect to the norm $|\cdot|_{p}$, that
is, the pair $E_{p}:=(D(A^{p}),|\cdot|_{p})$ is a Hilbert space. 
\item For $q\geq p\geq0$, let $j_{p,q}:E_{q}\rightarrow E_{p}$ be the
inclusion map. Then every inclusion map is continuous and has a dense
image. Then $\{ E_{p},j_{p,q}\}$ is a reduced projective system. 
\item A standard countable Hilbert space\[
E:=\lim_{\leftarrow}E_{p}=\bigcap_{p\geq0}E_{p}\]
constructed from the pair $(H,A)$ is a reflexive Fr{\'e}chet space.
We call $E$ a CH-space simply.
\item From (3), we have ${\displaystyle E^{*}=\lim_{\rightarrow}E_{p}^{*}}$
as a topological vector space, i.e. the strong topology on $E^{*}$
and the inductive topology on ${\displaystyle \lim_{\rightarrow}E_{p}^{*}}$
coincide.
\item Let $p\in\mathbf{R}_{\geq0}$ and $(x,y)_{-p}:=(A^{-p}x,A^{-p}y)_{0}$.
Then $(\cdot,\cdot)_{-p}$ is an inner product on $H$. 
\item For $p\geq0$, let $E_{-p}$ be the completion of $H$ with respect
to the norm $|\cdot|_{-p}$. For, $q\geq p\geq0$, we can consider
the inclusion map $i_{-q,-p}:E_{-p}\rightarrow E_{-q}$, and then
$\{ E_{-p},i_{-q,-p}\}$ is an inductive system. Moreover, $E_{-p}$
and $E_{p}^{*}$ are anti-linear isomorphic and isometric. Thus, from
(4), we have\[
E^{*}=\lim_{\rightarrow}E_{-p}=\bigcup_{p\geq0}E_{-p}.\]
 
\end{enumerate}
\end{defn}

Furthermore, we require for the operator $A$ that there exists $\alpha>0$
such that $A^{-\alpha}$ is a Hilbert-Schmidt class operator, namely\begin{equation}
\delta^{2}:=\sum_{j=1}^{\infty}\lambda_{j}^{-2\alpha}<\infty.\label{eq:Hilbert-Schmidt_inverse}\end{equation}
 From this condition, $E$ (resp. $E^{*}$) is a nuclear space. Thus
we can define the $\pi$-tensor topology $E\otimes_{\pi}E$ (resp.
$E^{*}\otimes_{\pi}E^{*}$) of $E$ (resp. $E^{*}$). If there is
no danger of confusion, we will use the notation $E\otimes E$ (resp.
$E^{*}\otimes E^{*}$) simply.

We denote the canonical bilinear form on $E^{*}\times E$ by $\left\langle \cdot,\cdot\right\rangle $.
We have the following natural relation between the canonical bilinear
form on $E^{*}\times E$ and the inner product on $H$ :\[
\left\langle f,g\right\rangle =(Jf,g)_{0}\]
for all $f\in H$ and $g\in E$. $Jf\in H$ is the complex conjugate
of $f\in H$. 

\begin{defn}
\label{thm:def_of_alternizer}Let $X$ be a Hilbert space, or a CH-space.
\begin{enumerate}
\item Let $g_{1}$, $\ldots$ , $g_{n}\in X$. We define the anti-symmetrization
$\mathcal{A}_{n}(g_{1}\otimes\ldots\otimes g_{n})$ of $g_{1}\otimes\ldots\otimes g_{n}\in H^{\otimes n}$
as follows.\[
\mathcal{A}_{n}(g_{1}\otimes\ldots\otimes g_{n}):=g_{1}\wedge\ldots\wedge g_{n}:=\frac{1}{n!}\sum_{\sigma\in\mathfrak{S}_{n}}\mathrm{sign}(\sigma)g_{\sigma(1)}\otimes\ldots\otimes g_{\sigma(n)},\]
where $\mathfrak{S}_{n}$ is the set of all permutations of $\{1,2,\ldots,n\}$
and $\mathrm{sign}(\sigma)$ is the signature of $\sigma\in\mathfrak{S}_{n}$. 
\item If $f\in X^{\otimes n}$ satisfies $\mathcal{A}_{n}(f)=f$, then we
call $f$ anti-symmetric. We denote the set of all anti-symmetric
elements of $X^{\otimes n}$ by $X^{\wedge n}$ and we call $X^{\wedge n}$
the $n$-th anti-symmetric tensor of $X$. If $X$ is a Hilbert space,
then $\mathcal{A}_{n}$ is a projection from $X^{\otimes n}$ to $X^{\wedge n}$
.
\item Let $X$ be a CH-space. For $F\in(X^{\otimes n})^{*}$ and $\sigma\in\mathfrak{S}_{n}$,
let $F^{\sigma}$ be an element of $(X^{\otimes n})^{*}$ satisfying\[
\left\langle F^{\sigma},g_{1}\otimes\ldots\otimes g_{n}\right\rangle :=\left\langle F,g_{\sigma^{-1}(1)}\otimes\ldots\otimes g_{\sigma^{-1}(n)}\right\rangle ,\quad g_{i}\in X.\]
Then we define the anti-symmetrization $\mathcal{A}_{n}(F)$ as follows.\[
\mathcal{A}_{n}(F):=\frac{1}{n!}\sum_{\sigma\in\mathfrak{S}_{n}}\mathrm{sign}(\sigma)F^{\sigma}.\]

\item If $F\in(X^{\otimes n})^{*}$ satisfies $\mathcal{A}_{n}(F)=F$, we
call $F$ anti-symmetric. We denote the set of all anti-symmetric
elements of $(X^{\otimes n})^{*}$ by $(X^{\wedge n})^{*}$.
\end{enumerate}
\end{defn}
From the above discussion, we obtain a Gelfand triple :\[
E\subset H\subset E^{*}.\]

\begin{lem}
Let $H$ be a Hilbert space. 
\begin{enumerate}
\item Let \[
(f_{1}\otimes\ldots\otimes f_{n},g_{1}\otimes\ldots\otimes g_{n})_{0}:=(f_{1},g_{1})_{0}\ldots(f_{n},g_{n})\]
for $f_{i}$, $g_{j}\in H$, $i,j=1,2,\ldots,n$. Then\[
(f_{1}\wedge\ldots\wedge f_{n},g_{1}\wedge\ldots\wedge g_{n})_{0}=\frac{1}{n!}\det\left((f_{i},g_{j})_{0}\right)_{1\leq i,j\leq n}.\]
Moreover $\mathcal{A}_{n}$ is a projection with respect to $(\cdot,\cdot)_{0}$.
\item For $f\in H^{\wedge n}$ and $g\in H^{\wedge m}$, we have\[
\left|f\wedge g\right|_{0}\leq\left|f\right|_{0}\left|g\right|_{0}.\]

\end{enumerate}
\end{lem}
Next, we define the Fermion Fock space and the second quantization
of a linear operator.

\begin{defn}
Let $H$ be a Hilbert space and $A$ be a linear operator on $H$.
\begin{enumerate}
\item Let\begin{gather*}
\Gamma(H):=\left\{ \sum_{n=0}^{\infty}\phi_{n}\ |\ \phi_{n}\in H^{\wedge n},\ \left\Vert \sum_{n=0}^{\infty}\phi_{n}\right\Vert _{0}^{2}:=\sum_{n=0}^{\infty}n!|\phi_{n}|_{0}^{\infty}<+\infty\right\} \\
\left(\sum_{n=0}^{\infty}\phi_{n},\sum_{n=0}^{\infty}\psi_{n}\right)_{0}:=\sum_{n\geq0}n!(\phi_{n},\psi_{n})_{0}\end{gather*}
Then we call $\Gamma(H)$ the Fermion Fock space. The Fermion Fock
space $\Gamma(H)$ is a Hilbert space with respect to the inner product
$\left\langle \left\langle \cdot,\cdot\right\rangle \right\rangle _{0}$.
Moreover let\begin{gather*}
\Gamma^{+}(H):=\left\{ \sum_{n=0}^{\infty}\phi_{2n}\,|\,\phi_{2n}\in H^{\wedge(2n)},\,\sum_{n=0}^{\infty}(2n)!\left|\phi_{2n}\right|_{0}^{2}<+\infty\right\} ,\\
\Gamma^{-}(H):=\left\{ \sum_{n=0}^{\infty}\phi_{2n+1}\,|\,\phi_{2n+1}\in H^{\wedge(2n+1)},\,\sum_{n=0}^{\infty}(2n+1)!\left|\phi_{2n+1}\right|_{0}^{2}<+\infty\right\} \end{gather*}
 Then we call $\Gamma^{+}(H)$ (resp. $\Gamma^{-}(H)$) the even part
of the Fermion Fock space ( resp. the odd part of the Fermion Fock
space).
\item We call \[
\Gamma(A):=\sum_{n=0}^{\infty}A^{\otimes n}\]
the second quantization of $A$. Let\[
\Gamma^{+}(A):=\Gamma(A)|\Gamma^{+}(H),\quad\Gamma^{-}(A):=\Gamma(A)|\Gamma^{-}(H).\]

\end{enumerate}
\end{defn}
\ 

\begin{defn}
\label{thm:def_of_CH-space_Gelfand_triple} Let $H$ be a complex
Hilbert space and $A$ be a self-adjoint operator on $H$ satisfying
the conditions (i) and (ii) in lemma \ref{thm:property_of_self-adj_op_A}
and \eqref{eq:Hilbert-Schmidt_inverse}. Then we can define a CH-space
$\mathcal{E}$ constructed from $(\Gamma(H),\Gamma(A))$ and we obtain
a Gelfand triple :\[
\mathcal{E}\subset\Gamma(H)\subset\mathcal{E}^{*}.\]
Moreover, let $\mathcal{E}_{+}$(resp. $\mathcal{E}_{-}$) be a CH-space
constructed from $(\Gamma^{+}(H),\Gamma^{+}(A))$ (resp. $(\Gamma^{-}(H),\Gamma^{-}(A))$)
and we obtain Gelfand triples :\[
\mathcal{E}_{+}\subset\Gamma^{+}(H)\subset\mathcal{E}_{+}^{*},\quad\mathcal{E}_{-}\subset\Gamma^{-}(H)\subset\mathcal{E}_{-}^{*}.\]
Then an element of $\mathcal{E}$ (or $\mathcal{E}_{+}$, $\mathcal{E}_{-}$)
is called a test (white noise) functional and an element of $\mathcal{E}^{*}$(or
$\mathcal{E}_{+}^{*}$, $\mathcal{E}_{-}^{*}$) is called a generalized
(white noise) functional.
\end{defn}
\begin{cor}
Let $\phi:=\sum_{n=0}^{\infty}\phi_{n}\in\Gamma(H)$, $\phi_{n}\in H^{\wedge n}$.
Then $\phi\in\mathcal{E}$ if and only if $\phi_{n}\in E^{\wedge n}$
for all $n\geq0$. Moreover, it holds that\[
\left\Vert \phi\right\Vert _{p}^{2}:=\left\Vert \Gamma(A)^{p}\phi\right\Vert _{0}^{2}=\sum_{n=0}^{\infty}n!|\phi_{n}|_{p}^{2}<+\infty\]
for all $p\geq0$. We can also show this statement in case of $\phi\in\Gamma^{+}(H)$
and $\phi\in\Gamma^{-}(H)$. 
\end{cor}
Let $H$ be a Hilbert space. Then $\zeta\wedge\eta=\eta\wedge\zeta$
for $\zeta$, $\eta\in H^{\wedge2}$. Thus we can define $\zeta^{\wedge n}$,
$n\geq0$ for all $\zeta\in H^{\wedge2}$. This shows that a pair
of Fermions behaves like a Boson. 

In order to discuss integral kernel operators, we define a contraction
of tensor product.

\begin{defn}
Let $H$ be a complex Hilbert space and $A$ be a self-adjoint operator
on $H$ satisfying the conditions (i) and (ii) in lemma \ref{thm:property_of_self-adj_op_A}
and \eqref{eq:Hilbert-Schmidt_inverse}. Let\[
e(\mathbf{i}):=e_{i_{1}}\otimes\ldots\otimes e_{i_{l}},\quad\mathbf{i}:=(i_{1},\ldots,i_{l})\in\mathbf{N}^{l}.\]

\end{defn}
\begin{enumerate}
\item For $F\in\left(E^{\otimes(l+m)}\right)^{*}$, let\[
|F|_{l,m;p,q}^{2}:=\sum_{\mathbf{i},\mathbf{j}}\left|\left\langle F,e(\mathbf{i})\otimes e(\mathbf{j})\right\rangle \right|^{2}\left|e(\mathbf{i})\right|_{p}^{2}\left|e(\mathbf{j})\right|_{q}^{2}\]
where $\mathbf{i}$ and $\mathbf{j}$ run the whole $\mathbf{N}^{l}$
and $\mathbf{N}^{m}$ respectively.
\item For $F\in\left(E^{\otimes(l+m)}\right)^{*}$ and $g\in E^{\otimes(m+n)}$,
we define a left contraction $F\otimes^{m}g\in\left(E^{\otimes(l+n)}\right)^{*}$
of $F$ and $g$ as follows.\[
F\otimes^{m}g:=\sum_{\mathbf{j},\mathbf{k}}\left(\sum_{\mathbf{i}}\left\langle F,e(\mathbf{i})\otimes e(\mathbf{j})\right\rangle \left\langle g,e(\mathbf{i})\otimes e(\mathbf{k})\right\rangle \right)e(\mathbf{j})\otimes e(\mathbf{k})\]
where $\mathbf{i}$, $\mathbf{j}$, and $\mathbf{k}$ run the whole
$\mathbf{N}^{m}$, $\mathbf{N}^{l}$, and $\mathbf{N}^{n}$ respectively.
Similarly, we define a right contraction $F\otimes_{m}g\in\left(E^{\otimes(l+n)}\right)^{*}$
of $F$ and $g$ as follows.\[
F\otimes_{m}g:=\sum_{\mathbf{j},\mathbf{k}}\left(\sum_{\mathbf{i}}\left\langle F,e(\mathbf{j})\otimes e(\mathbf{i})\right\rangle \left\langle g,e(\mathbf{k})\otimes e(\mathbf{i})\right\rangle \right)e(\mathbf{j})\otimes e(\mathbf{k})\]
where $\mathbf{i}$, $\mathbf{j}$, and $\mathbf{k}$ run the whole
$\mathbf{N}^{m}$, $\mathbf{N}^{l}$, and $\mathbf{N}^{n}$ respectively. 
\end{enumerate}
\begin{lem}
For $F\in(E^{\otimes(l+m)})^{*}$ and $g\in E^{\otimes(l+n)}$, put\begin{gather*}
F\wedge^{m}g:=\mathcal{A}_{l+n}(F\otimes^{m}g),\\
F\wedge_{m}g:=\mathcal{A}_{l+n}(F\otimes_{m}g).\end{gather*}
 Then $F\wedge^{m}g$ and $F\wedge_{m}g$ are elements of $(E^{\wedge(l+n)})^{*}$
and satisfy\[
F\wedge^{m}g=(-1)^{m(l+n)}F\wedge_{m}g.\]
 Thus the left contraction $F\wedge^{m}g$ coincides with the right
contraction $F\wedge_{m}g$ if $m$ is an even number.
\end{lem}
Before defining integral kernel operators, we need to mention continuity
of linear operators on locally convex spaces.

\begin{lem}
Let $X$ and $Y$ be locally convex spaces with seminorms $\{|\cdot|_{X,q}\}_{q\in Q}$
and $\{|\cdot|_{Y,p}\}_{p\in P}$ respectively. Let $\mathcal{L}(X,Y)$
be the set of all continuous linear operators from $X$ to $Y$. Then
a linear operator $V$ from $X$ to $Y$ is in $V\in\mathcal{L}(X,Y)$
if and only if, for any $p\in P$, there exist $q\in Q$ and $C>0$
such that\[
|Vx|_{Y,p}\leq C|x|_{X,q},\quad x\in X.\]

\end{lem}
Now we define an integral kernel operator.

\begin{prop}
[\bf Integral kernel operator]\label{thm:integral_kernel_op}Let
$\kappa\in((E^{\wedge2})^{\otimes(l+m)})^{*}$. For $\phi:=\sum_{n=0}^{\infty}\phi_{n}\in\mathcal{E}_{+}$,
$\phi_{n}\in E^{\wedge(2n)}$, let\[
\Xi_{l,m}(\kappa)\phi:=\sum_{n=0}^{\infty}\frac{(2n+2m)!}{(2n)!}\kappa\wedge_{2m}\phi_{m+n}.\]
Then $\Xi_{l,m}(\kappa)\in\mathcal{L}(\mathcal{E}_{+},\mathcal{E}_{+}^{*})$.
We call $\Xi_{l,m}(\kappa)$ an integral kernel operator with a kernel
distribution $\kappa$. This integral kernel operator satisfies the
following estimation :\[
\left\Vert \Xi_{l,m}(\kappa)\phi\right\Vert _{p}\leq\rho^{-\frac{r}{2}}((2l)^{2l}(2m)^{2m})^{\frac{1}{2}}\left(\frac{\rho^{-\frac{r}{2}}}{-re\log\rho}\right)^{l+m}|f|_{2l,2m;p,-q}\left\Vert \phi\right\Vert _{\mathrm{max}\{ p,q\}+r}\]
for any $\phi\in\mathcal{E}_{+}$. 
\end{prop}
Note that the following map\[
((E^{\wedge2})^{\otimes(l+m)})^{*}\ni\kappa\mapsto\Xi_{l,m}(\kappa)\in\mathcal{L}(\mathcal{E}_{+},\mathcal{E}_{+}^{*})\]
is not injective. We define\[
\mathcal{A}_{l,m}(\kappa):=\frac{1}{(2l)!(2m)!}\sum_{\sigma=(\sigma_{1},\sigma_{2})\in\mathfrak{S}_{2l}\times\mathfrak{S}_{2m}}\mathrm{sign}(\sigma_{1})\mathrm{sign}(\sigma_{2})\kappa^{\sigma},\]
where $\kappa^{\sigma}$ is defined in definition \ref{thm:def_of_alternizer}
(3). Put\[
(E^{\otimes(l+m)})_{\mathrm{alt}(l,m)}^{*}:=\{\kappa\in(E^{\otimes(l+m)})^{*}\,|\,\mathcal{A}_{l,m}(\kappa)=\kappa\,\}.\]
({}``alt'' stands for {}``alternative''.)

\begin{lem}
The map\[
((E^{\wedge2})^{\otimes(l+m)})_{\mathrm{alt}(2l,2m)}^{*}\ni\kappa\mapsto\Xi_{l,m}(\kappa)\in\mathcal{L}(\mathcal{E}_{+},\mathcal{E}_{+}^{*})\]
is injective. Moreover, for $\kappa\in((E^{\wedge2})^{\otimes(l+m)})_{\mathrm{alt}(2l,2m)}^{*}$
and $\kappa\in((E^{\wedge2})^{\otimes(l'+m')})_{\mathrm{alt}(2l',2m')}^{*}$,
if $\Xi_{l,m}(\kappa)=\Xi_{l',m'}(\kappa')$, then $l=l'$, $m=m'$,
and $\mathcal{A}_{2l,2m}(\kappa)=\mathcal{A}_{2l,2m}(\kappa')$.
\end{lem}
\begin{prop}
[\bf Fock expansion]\label{thm:operator_Fock_expansion_even_part}
For any $\Xi\in\mathcal{L}(\mathcal{E}_{+},\mathcal{E}_{+}^{*})$,
there exists a unique $\{\kappa_{l,m}\}_{l,m=0}^{\infty}$, $\kappa_{l,m}\in((E^{\wedge2})^{\otimes(l+m)})_{\mathrm{alt}(2l,2m)}^{*}$
such that \begin{equation}
\Xi\phi=\sum_{l,m=0}^{\infty}\Xi_{l,m}(\kappa_{l,m})\phi,\quad\phi\in\mathcal{E}_{+}\label{eq:Fock_expansion}\end{equation}
where the right hand side of \eqref{eq:Fock_expansion} converges in
$\mathcal{E}_{+}^{*}$.

If $\Xi\in\mathcal{L}(\mathcal{E}_{+},\mathcal{E}_{+})$, then\[
\kappa_{l,m}\in E^{\wedge(2l)}\otimes\left(E^{\wedge(2m)}\right)^{*},\quad l,m\geq0\]
and the right hand side of \eqref{eq:Fock_expansion} converges in $\mathcal{E}_{+}$.
\end{prop}
Now we remark

\begin{lem}
$\Xi_{l,m}(\kappa_{l,m})\in\mathcal{L}(\mathcal{E}_{+},\mathcal{E}_{+}^{*})$
is extended to an element of $\mathcal{L}(\Gamma^{+}(H),\mathcal{E}_{+}^{*})$
if and only if $\kappa_{l,m}$ is in $(E^{\wedge(2l)})^{*}\otimes H^{\wedge(2m)}$.
\end{lem}
Now we extend results of proposition \ref{thm:operator_Fock_expansion_even_part}
to all elements of $\mathcal{L}(\mathcal{E},\mathcal{E}^{*})$. In
order to make the white noise calculus for the Fermion system, we
mention properties for creation and annihilation operators for the
Fermion system.

\begin{defn}
[\bf Creation and  annihilation operator]\ $\,$
\end{defn}
\begin{enumerate}
\item For $f\in E^{*}$, let\begin{gather*}
a(f):\mathcal{E}\ni\phi=(\phi_{n})_{n\in\mathbf{Z}\geq0}\mapsto a(f)\phi\in\mathcal{E},\\
a(f)\phi_{0}=0,\\
a(f)\phi_{n}:=n\mathcal{A}_{n-1}(f\otimes_{1}\phi_{n}),\quad n\geq1.\end{gather*}
$a(f)$ is a map from $\mathcal{E}$ to $\mathcal{E}$ and is a continuous
map. (Continuity of $a(f)$ is seen in the following lemma.) We call
$a(f)$ an annihilation operator.
\item For $f\in E^{*}$, we define a creation operator $a^{\dagger}(f)\in\mathcal{L}(\mathcal{E}^{*},\mathcal{E}^{*})$
as follows :\begin{gather*}
a^{\dagger}(f):\mathcal{E}^{*}\ni\phi=(\phi_{n})_{n\in\mathbf{Z}\geq0}\mapsto a^{\dagger}(f)\phi\in\mathcal{E}^{*},\\
a^{\dagger}(f)\phi_{n}:=\mathcal{A}_{n+1}(f\otimes\phi_{n}),\quad n\geq0.\end{gather*}
(Continuity of $a^{\dagger}(f)$ is seen in the following lemma.)
\end{enumerate}
\begin{lem}
\label{lem:basic property of creation annihilation op}Let\[
a_{(l,m)}(f):=\left\{ \begin{array}{ccc}
a^{\dagger}(f), &  & \mathrm{if}\ (l,m)=(1,0)\\
a(f), &  & \mathrm{if}\ (l,m)=(0,1).\end{array}\right.\]
and $p,\  q\in\mathbf{R}$, $r>0$, and $f\in E^{*}$. Then\[
\left\Vert a_{(l,m)}(f)\phi\right\Vert _{p}\leq\left(\frac{\rho^{-2r}}{-2re\log\rho}\right)^{\frac{1}{2}}|f|_{m,l;-(q+r),p}\left\Vert \phi\right\Vert _{\mathrm{max}\{ p,q\}+r},\quad\phi\in\mathcal{E}.\]
Thus we have the following properties (1)--(3). Let $\sigma$ be $+$,
$-$, or a blank.
\begin{enumerate}
\item $a(f)|_{\mathcal{E}_{\sigma}}\in\mathcal{L}(\mathcal{E}_{\sigma},\mathcal{E}_{-\sigma})$
for $f\in E^{*}$,
\item $a^{\dagger}(f)|_{\mathcal{E}_{\sigma}}\in\mathcal{L}(\mathcal{E}_{\sigma},\mathcal{E}_{-\sigma})$
for $f\in E$,
\item $a^{\dagger}(f)|_{\mathcal{E}_{\sigma}^{*}}=a(f)^{*}|_{\mathcal{E}_{\sigma}^{*}}\in\mathcal{L}(\mathcal{E}_{\sigma}^{*},\mathcal{E}_{-\sigma}^{*})$
for $f\in E^{*}$,
\end{enumerate}
\end{lem}
Creation and annihilation operators satisfy the following commutation
relation, called canonical anti-commutation relations. 

\begin{lem}
\label{thm:anti-commutation-relations}For $f\in E$ and $g\in E^{*}$,
we have\[
\{ a^{\dagger}(f),a(g)\}\phi=\left\langle g,f\right\rangle \phi\]
for $\phi\in\mathcal{E}_{\sigma}$. ($\sigma$ is $+$, $-$, or a
blank. ) Moreover,\[
(a^{\dagger}(f)+a(Jf))^{2}\phi=\phi\]
for $\phi\in\mathcal{E}_{\sigma}$ and $f\in E$ with $(f,f)_{K}=1$.
Here $Jf\in E$ is the complex conjugate of $f\in E$. 
\end{lem}
Put \[
d\Gamma(A)^{(n)}:=\sum_{i=1}^{n}1^{\otimes(i-1)}\otimes A\otimes1^{\otimes(n-i)}\]
 for a linear operator $A\in\mathcal{L}(E,E^{*})$. 

\begin{lem}
$\,$
\begin{enumerate}
\item For $f$, $g\in E^{*}$, we have\begin{gather}
a^{\dagger}(f)a^{\dagger}(g)|_{\mathcal{E}_{+}}=\Xi_{1,0}(f\wedge g),\label{eq:creation_creation_equal_2particle_creation}\\
a(f)a(g)|_{\mathcal{E}_{+}}=\Xi_{0,1}(f\wedge g).\label{eq:annihilation_annihilation_equal_2particle_annihilation}\end{gather}

\item Let $(f\otimes g)h:=\left\langle g,h\right\rangle f$ for $f$, $g\in E^{*}$,
and $h\in E$. Then we have\begin{equation}
a^{\dagger}(f)a(g)|_{\mathcal{E}_{+}}=\Xi_{1,1}((1_{2}\otimes d\Gamma(f\otimes g)^{(2)})^{*}\tau)\label{eq:creation_annihilation_equal_second_quantization}\end{equation}
where $\tau\in(E^{\wedge2})\otimes(E^{\wedge2})^{*}$ is defined by
$\tau(\zeta,\eta):=\left\langle \zeta,\eta\right\rangle $ for all
$\zeta\in(E^{\wedge2})^{*}$ and $\eta\in E^{\wedge2}$. 
\end{enumerate}
\end{lem}
Let $\Xi_{l,m}(\kappa)\in\mathcal{L}(\mathcal{E}_{+},\mathcal{E}_{+}^{*})$
be an integral kernel operator with a kernel distribution $\kappa$
and $f,\, g$ be elements of $E$. Put\[
W(f):=a^{\dagger}(f)+a(Jf)\]
for $f\in E$ satisfying $(f,f)_{K}=1$. Note that $W(f)$ is in $\mathcal{L}(\mathcal{E},\mathcal{E})$
and $\mathcal{L}(\mathcal{E}^{*},\mathcal{E}^{*})$.(See Lemma \ref{lem:B(x) is in L(E_E) and L(E^*_E^*)})
Then we also call all operators\[
\Xi_{l,m}(\kappa)W(f),\quad W(f)\Xi_{l,m}(\kappa),\quad W(f)\Xi_{l,m}(\kappa)W(f)\]
integral kernel operators for the sake of convenience. 

\begin{prop}
Every $\Xi\in\mathcal{L}(\mathcal{E},\mathcal{E}^{*})$ is realized
as a series of integral kernel operators.
\end{prop}
\begin{proof}
See Theorem 5.5 of \cite{Shimada 2}. 
\end{proof}

\section{Bogoliubov automorphisms and implementability conditions}

We apply the white noise calculus to the representation theory of
Bogoliubov transformations. 

We will begin by defining the canonical anti-commutation relations
algebra (referred to as CAR algebra henceforth).

Let $K$ be a Hilbert space and $(f,g)_{K}$ be the inner product
of $f\in K$ and $g\in K$. Let $\Gamma$ be an anti-unitary involution
on $K$. Let $\mathfrak{A}(K,\Gamma)$ be a $C^{*}$-algebra generated
by $B(f)$ $(f\in K)$ satisfying the following commutation relations
:\[
\{ B(f)^{\dagger},B(g)\}=(f,g)_{K}1,\quad B(f)^{\dagger}=B(\Gamma f)\]
for any $f$, $g\in K$. ($\dagger$ is the star operation of $C^{*}$-algebra
$\mathfrak{A}(K,\Gamma)$.) We call $\mathfrak{A}(K,\Gamma)$ the
(self-dual) CAR algebra.

We define a positive energy Fock representation of the CAR algebra. 

Let $\mathfrak{\mathfrak{h}}$ be a self-adjoint operator on $K$
and $P$ be a projection on $K$ commuting with $\mathfrak{h}$ and
satisfying $\Gamma P\Gamma=1-P$. Assume that $A:=\mathfrak{h}P$
has properties (i) and (ii) of definition \ref{thm:property_of_self-adj_op_A}. 

In order to discuss the positive energy Fock representation of the
CAR algebra $\mathfrak{A}(K,\Gamma)$, we divide $K$ into the positive
energy part $P_{+}K$ and the negative energy part $P_{-}K$. Let
$H:=P_{+}K$ and\[
(f,g)_{0}:=(f,g)_{K},\quad f,\, g\in H\]
be the inner product on $H$. Let $J$ be an anti-linear isomorphism
on $K$ commuting with $P$, i.e., $J$ gives a complex structure
of $H$.

\begin{defn}
Let $\pi_{P_{+}}$ be the map {*}-homomorphism $\pi_{P}$ from $\mathfrak{A}(K,\Gamma)$
to $\mathcal{L}(\Gamma(H))$ satisfying\[
\pi_{P_{+}}(B(f)):=a^{\dagger}(P_{+}f)+a(JP_{+}\Gamma f)\]
for any $f\in K$. We call $\pi_{P_{+}}$ a positive energy Fock representation
of the CAR algebra with respect to $P_{+}$. (See (2.15) of \cite{Araki_Contemp_math}.)
\end{defn}

\begin{defn}
$\,$
\begin{enumerate}
\item Let $U\in\mathcal{L}(K)$ be a unitary operator commuting with $\Gamma$.
Then we call $U$ a Bogoliubov transformation for the pair $(K,\Gamma)$.
Then we use $O(K,\Gamma)$ to denote all Bogoliubov transformations. 
\item Put\[
o(K,\Gamma):=\{ X\in\mathcal{L}(K)\,|\,\exp(\sqrt{-1}tX)\in O(K,\Gamma)\,\,\mathrm{for}\,\mathrm{all}\,\, t\in\mathbf{R}\}.\]
Then $X\in o(K,\Gamma)$ if and only if $X\in\mathcal{L}(K)$ satisfies
$X^{\dagger}=X$ and $\Gamma X\Gamma=-X$, where $X^{\dagger}$ is
defined by $(X^{\dagger}f,g)_{K}=(f,Xg)_{K}$ for all $f$, $g\in K$.
\item $o(K,\Gamma)$ is a Lie algebra with respect to $\sqrt{-1}[\cdot,\cdot]$.
Let $o_{\mathrm{fin}}(K,\Gamma)$ be the set of all $A\in o(K,\Gamma)$
with finite rank. $o_{\mathrm{fin}}(K,\Gamma)$ is a Lie subalgebra
of $o(K,\Gamma)$.
\item Let $E$ be a CH-space constructed from $(H,A)$. (See Definition
\ref{thm:def_of_CH-space_Gelfand_triple}.) Let\begin{gather*}
o(K,\Gamma;E):=\{ X\in o(K,\Gamma)\,|\, X(E\oplus\Gamma E)\subset(E\oplus\Gamma E)\},\\
o_{\mathrm{fin}}(K,\Gamma;E):=o(K,\Gamma;E)\cap o_{\mathrm{fin}}(K,\Gamma).\end{gather*}
Then $o(K,\Gamma;E)$ is a subalgebra of $o(K,\Gamma)$ and $o_{\mathrm{fin}}(K,\Gamma;E)$
is a subalgebra of $o_{\mathrm{fin}}(K,\Gamma)$.
\item Consider a map\begin{gather*}
\mathrm{ad}(X):o_{\mathrm{fin}}(K,\Gamma)\rightarrow o_{\mathrm{fin}}(K,\Gamma),\\
\mathrm{ad}(X)(Y):=[X,Y],\quad Y\in o_{\mathrm{fin}}(K,\Gamma).\end{gather*}
Then {}``$\mathrm{ad}$'' is a Lie algebra homomorphism from $o(K,\Gamma)$
to $\mathcal{L}(o_{\mathrm{fin}}(K,\Gamma))$.
\item Let $Y$ be a finite rank operator on $K$ satisfying $Yh=\sum_{i=1}^{n}(g_{i},h)_{K}f_{i}$
for any $h\in K$. Then\[
q(Y):=\frac{1}{2}\sum_{i=1}^{n}B(f_{i})B(\Gamma g_{i}).\]
$q(Y)$ does not depend on a choice of $f_{i}$, $g_{i}\in K$, namely
$q$ is well-defined as a map from finite rank operators on $K$ to
CAR algebra $\mathfrak{A}(K,\Gamma)$. (See Theorem 4.4. of \cite{Araki_Contemp_math}.) 
\item $q$ is a Lie algebra homomorphism from $o_{\mathrm{fin}}(K,\Gamma)$
to $\mathfrak{A}(K,\Gamma)$. {}``$q$'' stands for {}``quantization''
of Lie algebra $o_{\mathrm{fin}}(K,\Gamma)$. Thus $q_{P_{+}}:=\pi_{P_{+}}\circ q$
is a Lie algebra homomorphism from $o_{\mathrm{fin}}(K,\Gamma)$ to
$\mathcal{L}(\Gamma(H))$. In other words, $q_{P_{+}}$ is a Lie algebra
representation of $o_{\mathrm{fin}}(K,\Gamma)$ on the Fermion Fock
space $\Gamma(H)$.
\end{enumerate}
\end{defn}
Let $\mathcal{E}$ be a CH-space constructed from ($\Gamma(H)$,$\Gamma(A)$).
We remark that

\begin{lem}
\label{lem:B(x) is in L(E_E) and L(E^*_E^*)}We use the same notation
$\pi_{P_{+}}(B(x)))$ (resp. $q_{P_{+}}(X)$) to denote the extension
of $\pi_{P_{+}}(B(x)))$ (resp. $q_{P_{+}}(X)$) to $\Gamma(H)$ or
$\mathcal{E}^{*}$. 
\begin{enumerate}
\item $\pi_{P_{+}}(B(x)))$ is in $\mathcal{L}(\Gamma(H))$ for $x\in K$,
and $\pi_{P_{+}}(B(x)))$ is in $\mathcal{L}(\mathcal{E},\mathcal{E})$
and $\mathcal{L}(\mathcal{E}^{*},\mathcal{E}^{*})$ for $x\in E\oplus\Gamma E$.
\item $q_{P_{+}}(X)$ is in $\mathcal{L}(\mathcal{E},\Gamma(H))$ and $\mathcal{L}(\Gamma(H),\mathcal{E}^{*})$
for $X\in o_{\mathrm{fin}}(K,\Gamma)$, and $q_{P_{+}}(X)$ is in
$\mathcal{L}(\mathcal{E},\mathcal{E})$ and $\mathcal{L}(\mathcal{E}^{*},\mathcal{E}^{*})$
for $X\in o_{\mathrm{fin}}(K,\Gamma;E)$.
\end{enumerate}
\end{lem}
\begin{proof}
(1) $\pi_{P_{+}}(B(x)))\in\mathcal{L}(\mathcal{E},\mathcal{E})$ is
immediate from Lemma \ref{lem:basic property of creation annihilation op},
and $\pi_{P_{+}}(B(x)))\in\mathcal{L}(\mathcal{E}^{*},\mathcal{E}^{*})$
is immediate from\[
\pi_{P_{+}}(B(x)))=a^{\dagger}(P_{+}x)+(a^{\dagger}(JP_{+}\Gamma x)|_{\mathcal{E}})^{*}\]
on $\mathcal{E}^{*}$. Due to (1), we can check (2) easily.
\end{proof}
Now we give the definition of implementability of Bogoliubov automorphisms
at Lie algebraic level, that is,

\begin{defn}
\label{thm:definition_of_implementability}$\,$
\begin{enumerate}
\item Let $X\in o(K,\Gamma;E)$. We call $\mathrm{ad}(X)$ implementable
as $\mathcal{L}(\mathcal{E},\mathcal{E}^{*})$ if there exists implementer
$M_{X}\in\mathcal{L}(\mathcal{E},\mathcal{E}^{*})$ satisfying\begin{equation}
M_{X}q_{P_{+}}(Y)\phi-q_{P_{+}}(Y)M_{X}\phi=q_{P_{+}}(\mathrm{ad}(X)(Y))\phi\label{eq:implements_automorph(differential_ver.)}\end{equation}
for all $Y\in o_{\mathrm{fin}}(K,\Gamma;E)$ and $\phi\in\mathcal{E}$.
We denote the left hand side of \eqref{eq:implements_automorph(differential_ver.)}
by $[M_{X},q_{P_{+}}(Y)]$. (Note that the use of the notation of
the commutator {[} , {]} is only for convenience.)
\item Let $X\in o(K,\Gamma;E)$. We call $\mathrm{ad}(X)$ implementable
as $\mathcal{L}(\mathcal{E},\mathcal{E})$ if there exists $M_{X}\in\mathcal{L}(\mathcal{E},\mathcal{E})$
satisfying \eqref{eq:implements_automorph(differential_ver.)} for all
$Y\in o_{\mathrm{fin}}(K,\Gamma;E)$.
\item Let $X\in o(K,\Gamma)$. We call $\mathrm{ad}(X)$ implementable as
$\mathcal{L}(\Gamma(H),\mathcal{E}^{*})$ if there exists $M_{X}\in\mathcal{L}(\Gamma(H),\mathcal{E}^{*})$
satisfying \eqref{eq:implements_automorph(differential_ver.)} for all
$Y\in o_{\mathrm{fin}}(K,\Gamma)$.
\end{enumerate}
\end{defn}
Remark that $\mathrm{ad}(X)$ is automatically implementable as $\mathcal{L}(\mathcal{E},\mathcal{E}^{*})$
when $\mathrm{ad}(X)$ is implementable as $\mathcal{L}(\mathcal{E},\mathcal{E})$.
Moreover, $\mathrm{ad}(X)$ is also implementable as $\mathcal{L}(\mathcal{E},\mathcal{E}^{*})$
if $\mathrm{ad}(X)$ is implementable as $\mathcal{L}(\Gamma(H),\mathcal{E}^{*})$. 

Our purpose of this section is to give necessary and sufficient conditions
for existence of $M_{X}$ given in definition \ref{thm:definition_of_implementability}
in terms of $X\in o(K,\Gamma)$. We now say our result of this section.

\begin{thm}
\label{thm:implementability_condition}$\,$
\begin{enumerate}
\item Let $X\in o(K,\Gamma;E)$. There exists $M_{X}\in\mathcal{L}(\mathcal{E},\mathcal{E}^{*})$
satisfying \eqref{eq:implements_automorph(differential_ver.)} for all
$A\in o_{\mathrm{fin}}(K,\Gamma;E)$ if and only if there exists $p>0$
such that\begin{equation}
\sum_{n\in\mathbf{N}}\lambda_{n}^{-2p}\left|P_{+}XP_{-}\Gamma e_{n}\right|_{-p}^{2}<\infty.\label{eq:implement_cond(test_func_to_gen_func)}\end{equation}

\item Let $X\in o(K,\Gamma;E)$. There exists $M_{X}\in\mathcal{L}(\mathcal{E},\mathcal{E})$
satisfying \eqref{eq:implements_automorph(differential_ver.)} for all
$A\in o_{\mathrm{fin}}(K,\Gamma;E)$ if and only if $X\in o(K,\Gamma;E)$
satisfies\begin{equation}
\sum_{n\in\mathbf{N}}\lambda_{n}^{2p}\left|P_{+}XP_{-}\Gamma e_{n}\right|_{p}^{2}<\infty\label{eq:implement_cond(test_func_sp_preserving)}\end{equation}
for any $p>0$.
\item Let $X\in o(K,\Gamma)$. There exists $M_{X}\in\mathcal{L}(\Gamma(H),\mathcal{E}^{*})$
satisfying \eqref{eq:implements_automorph(differential_ver.)} for all
$A\in o_{\mathrm{fin}}(K,\Gamma)$ if and only if $P_{+}XP_{-}$ is
a Hilbert-Schmidt class operator on $K$, that is, \begin{equation}
\sum_{n\in\mathbf{N}}\left|P_{+}XP_{-}\Gamma e_{n}\right|_{0}^{2}<\infty.\label{eq:implemet_cond(H.S.cond)}\end{equation}

\end{enumerate}
In (1), (2), and (3), all $M_{X}$ are uniquely determined except
for constant numbers.

\end{thm}
We have some facts that should be noticed before we prove theorem
\ref{thm:implementability_condition}.

First, we can also show \eqref{eq:implements_automorph(differential_ver.)}
for $Y\in\mathbf{C}\otimes o_{\mathrm{fin}}(K,\Gamma)$ by linearity
of $q$. Now let\begin{gather*}
H_{1}(x,y)z:=\frac{1}{2}\left\{ \left(y,z\right)_{K}x+\left(x,z\right)_{K}y-\left(\Gamma y,z\right)_{K}\Gamma x-\left(\Gamma x,z\right)_{K}\Gamma y\right\} ,\\
H_{2}(x,y)z:=\frac{\sqrt{-1}}{2}\left\{ \left(x,z\right)_{K}y-\left(y,z\right)_{K}x+\left(\Gamma x,z\right)_{K}\Gamma y-\left(\Gamma y,z\right)_{K}\Gamma x\right\} \end{gather*}
for $x$, $y\in K$ and $z\in K$. Then $H_{1}(x,y)$, $H_{2}(x,y)\in o_{\mathrm{fin}}(K,\Gamma)$.
If\begin{equation}
Y=\frac{1}{2}\left(H_{1}(x,\Gamma y)+\sqrt{-1}H_{2}(x,\Gamma y)\right)\in\mathbf{C}\otimes o_{\mathrm{fin}}(K,\Gamma)\label{eq:A_equal_H_1+sqrt(-1)H_2}\end{equation}
for $x$, $y\in K$, then \eqref{eq:implements_automorph(differential_ver.)}
implies\begin{equation}
[M_{X},\pi_{P_{+}}(B(x)B(y))]=\pi_{P_{+}}(B(Xx)B(y)+B(x)B(Xy))\label{eq:implements_automorph(differential_ver)-No2}\end{equation}
for $x$, $y\in K$.

Secondly, we have the following lemma.

\begin{lem}
\label{thm:even->even_odd->odd}Assume that $M_{X}$ is a linear operator
satisfying \eqref{eq:implements_automorph(differential_ver.)}. Then
$M_{X}$ satisfies\[
M_{X}\mathcal{E}_{\sigma}^{*}\subset\mathcal{E}_{\sigma}^{*},\quad\sigma\in\{+,-\}.\]

\end{lem}
This lemma is easily checked. Let $M_{X}^{\sigma}:=M_{X}|_{\mathcal{E}_{\sigma}^{*}}$,
$\sigma\in\{+,-\}$. From the above two facts, \eqref{eq:implements_automorph(differential_ver)-No2}
is decomposed into the even part and the odd part, i.e., we have\begin{equation}
[M_{X}^{\sigma},q_{P_{+}}(A)]|_{\mathcal{E}_{\sigma}}=q_{P_{+}}(\mathrm{ad}(X)(A))|_{\mathcal{E}_{\sigma}}\label{eq:implements_automorph(diff_ver)-odd_and_even_case}\end{equation}
for $x$, $y\in K$ and $\sigma\in\{+,-\}$. Hence theorem \ref{thm:implementability_condition}
are equivalent to the following lemma.

\begin{lem}
\label{thm:implementability_ condition(odd_even_decomp_ver)}$\,$
\begin{enumerate}
\item Let $X\in o(K,\Gamma;E)$. There exists $M_{X}^{\sigma}\in\mathcal{L}(\mathcal{E}_{\sigma},\mathcal{E}_{\sigma}^{*})$
satisfying \eqref{eq:implements_automorph(diff_ver)-odd_and_even_case}
for all $A\in o_{\mathrm{fin}}(K,\Gamma;E)$ if and only if there
exists $p>0$ such that\[
\sum_{n\in\mathbf{N}}\lambda_{n}^{-2p}\left|P_{+}XP_{-}\Gamma e_{n}\right|_{-p}^{2}<\infty.\]

\item Let $X\in o(K,\Gamma;E)$. There exists $M_{X}\in\mathcal{L}(\mathcal{E}_{\sigma},\mathcal{E}_{\sigma})$
satisfying \eqref{eq:implements_automorph(diff_ver)-odd_and_even_case}
for all $A\in o_{\mathrm{fin}}(K,\Gamma;E)$ if and only if $X\in o(K,\Gamma;E)$
satisfies\[
\sum_{n\in\mathbf{N}}\lambda_{n}^{2p}\left|P_{+}XP_{-}\Gamma e_{n}\right|_{p}^{2}<\infty\]
for any $p>0$.
\item Let $X\in o(K,\Gamma)$. There exists $M_{X}^{\sigma}\in\mathcal{L}(\Gamma^{\sigma}(H),\mathcal{E}_{\sigma}^{*})$
satisfying \eqref{eq:implements_automorph(diff_ver)-odd_and_even_case}
for all $A\in o_{\mathrm{fin}}(K,\Gamma)$ if and only if $P_{+}XP_{-}$
is a Hilbert-Schmidt class operator on $K$, that is,\[
\sum_{n\in\mathbf{N}}\left|P_{+}XP_{-}\Gamma e_{n}\right|_{0}^{2}<\infty.\]

\end{enumerate}
\end{lem}
Thirdly, we describe the composition of two integral kernel operators.
We define an element $S_{m-k}^{l}\,_{m'}^{l'-k}(\kappa\circ_{k}\lambda)$
of $\left(E_{+}^{\otimes(l+l'+m+m'-2k)}\right)^{*}$ as follows :\begin{align*}
S_{m-k}^{l}\,_{m'}^{l'-k}(\kappa\circ_{k}\lambda):=\sum_{\mathbf{i},\mathbf{j},\mathbf{i}',\mathbf{j}'}\sum_{\mathbf{h}} & \left\langle \kappa,e(\mathbf{i})\otimes e(\mathbf{j})\otimes e(\mathbf{h})\right\rangle \\
 & \times\left\langle \lambda,e(\mathbf{h})\otimes e(\mathbf{i}')\otimes e(\mathbf{j}')\right\rangle e(\mathbf{i})\otimes e(\mathbf{i}')\otimes e(\mathbf{j})\otimes e(\mathbf{j}'),\end{align*}
where $\mathbf{i}$, $\mathbf{j}$, $\mathbf{i}'$, $\mathbf{j}'$
and $\mathbf{h}$ run over the whole of $\mathbf{N}^{l}$, $\mathbf{N}^{m-k}$,
$\mathbf{N}^{l'-k}$, $\mathbf{N}^{m'}$ and $\mathbf{N}^{k}$ respectively.
Then we have\begin{equation}
\Xi_{l,m}(\kappa)\Xi_{l',m'}(\lambda)=\sum_{k=0}^{\mathrm{min}\{ m,l'\}}k!\left(\begin{array}{c}
m\\
k\end{array}\right)\left(\begin{array}{c}
l'\\
k\end{array}\right)\Xi_{l+l'-k,m+m'-k}(S_{m-k}^{l}\,_{m'}^{l'-k}(\kappa\circ_{k}\lambda)).\label{eq:composition_of_two_int_ker_ops}\end{equation}
for $\Xi_{l,m}(\kappa)\in\mathcal{L}(\mathcal{E}_{+},\mathcal{E}_{+}^{*})$
and $\Xi_{l',m'}(\lambda)\in\mathcal{L}(\mathcal{E}_{+},\mathcal{E}_{+})$.
As for well-definedness of $S_{m-k}^{l}\,_{m'}^{l'-k}(\kappa\circ_{k}\lambda)$,
see Definition 2.13 of \cite{Shimada}. 

\begin{rem}
We have the following equivalent conditions of implementability of
$\mathrm{ad}(X)$. 
\begin{enumerate}
\item Let $X\in o(K,\Gamma;E)$. $\mathrm{ad}(X)$ is implementable as $\mathcal{L}(\mathcal{E},\mathcal{E}^{*})$
if and only if there exists $M_{X}\in\mathcal{L}(\mathcal{E},\mathcal{E}^{*})$
satisfying\begin{equation}
M_{X}\pi_{P_{+}}(B(x))\phi-\pi_{P_{+}}(B(x))M_{X}\phi=\pi_{P_{+}}(B(Xx))\phi\label{eq:implements_auto([M_X,B(f)]_type)}\end{equation}
for all $x\in E\oplus\Gamma E$ and $\phi\in\mathcal{E}$.
\item Let $X\in o(K,\Gamma;E)$. $\mathrm{ad}(X)$ is implementable as $\mathcal{L}(\mathcal{E},\mathcal{E})$
if and only if there exists $M_{X}\in\mathcal{L}(\mathcal{E},\mathcal{E})$
satisfying \eqref{eq:implements_auto([M_X,B(f)]_type)} for all $x\in E\oplus\Gamma E$
and $\phi\in\mathcal{E}$.
\item Let $X\in o(K,\Gamma)$. $\mathrm{ad}(X)$ is implementable as $\mathcal{L}(\Gamma(H),\mathcal{E}^{*})$
if and only if there exists $M_{X}\in\mathcal{L}(\Gamma(H),\mathcal{E}^{*})$
satisfying \eqref{eq:implements_auto([M_X,B(f)]_type)} for all $x\in K$
and $\phi\in\Gamma(H)$.
\end{enumerate}
\end{rem}
\begin{proof}
It suffices to show (1) and (2). (In case of (3) we have only to replace
the space $E\oplus\Gamma E$ in the following proof with $K$.) Suppose
\eqref{eq:implements_auto([M_X,B(f)]_type)}. Then \eqref{eq:implements_automorph(differential_ver)-No2}
follows from direct computation. Conversely, we assume \eqref{eq:implements_automorph(differential_ver)-No2}
and prove \eqref{eq:implements_auto([M_X,B(f)]_type)}. Let\[
M'_{X}:=2\pi_{P_{+}}(B(x))M_{X}\pi_{P_{+}}(B(x))-2\pi_{P_{+}}(B(x)B(Xx))\]
for $x$, $y\in E\oplus\Gamma E$ satisfy $\Gamma x=x$ and $|x|_{K}=1$.
Then $M'_{X}$ is independent of the choice of $x\in E\oplus\Gamma E$.
In fact,\begin{align*}
\pi_{P_{+}} & (B(x))M_{X}\pi_{P_{+}}(B(x))-\pi_{P_{+}}(B(x)B(Xx))\\
 & =\pi_{P_{+}}(B(x))M_{X}\pi_{P_{+}}(B(x))\cdot2\pi_{P_{+}}(B(y))^{2}-\pi_{P_{+}}(B(x)B(Xx))\\
 & =2\pi_{P_{+}}(B(x))\{\pi_{P_{+}}(B(x)B(y))M_{X}\\
 & \qquad+\pi_{P_{+}}(B(Xx)B(y))+\pi_{P_{+}}(B(x)B(Xy))\}\pi_{P_{+}}(B(y))-\pi_{P_{+}}(B(x)B(Xx))\\
 & =\pi_{P_{+}}(B(y))M_{X}\pi_{P_{+}}(B(y))+\pi_{P_{+}}(B(Xy)B(y)),\\
 & =\pi_{P_{+}}(B(y))M_{X}\pi_{P_{+}}(B(y))-\pi_{P_{+}}(B(y)B(Xy)).\end{align*}
for $x$, $y\in E\oplus\Gamma E$ satisfying $\Gamma x=x$, $\Gamma y=y$
and $|x|_{K}=|y|_{K}=1$. (The last equation of the above calculation
is led by $(y,Xy)_{K}=0$ for $X\in o(K,\Gamma)$ and $y\in E\oplus\Gamma E$
with $\Gamma y=y$.) From the definition of $M'_{X}$, we have\begin{gather}
M'_{X}\pi_{P_{+}}(B(x))=\pi_{P_{+}}(B(x))M_{X}+\pi_{P_{+}}(B(Xx)),\label{eq:relation of M_X and M'_X(1)}\\
\pi_{P_{+}}(B(x))M'_{X}=M_{X}\pi_{P_{+}}(B(x))-\pi_{P_{+}}(B(Xx))\label{eq:relation of M_X and M'_X(2)}\end{gather}
for any $x\in E\oplus\Gamma E$ satisfying $\Gamma x=x$ and $|x|_{K}=1$.
(By linearity of $B(\cdot)$, \eqref{eq:relation of M_X and M'_X(1)}
and \eqref{eq:relation of M_X and M'_X(2)} hold for any $x\in E\oplus\Gamma E$.)
Thus\begin{gather*}
M'_{X}\pi_{P_{+}}(B(x)B(y))=\pi_{P_{+}}(B(x))M_{X}\pi_{P_{+}}(B(y))+\pi_{P_{+}}(B(Xx)B(y)),\\
\pi_{P_{+}}(B(x)B(y))M'_{X}=\pi_{P_{+}}(B(x))M_{X}\pi_{P_{+}}(B(y))-\pi_{P_{+}}(B(Xx)B(y))\end{gather*}
for all $x$, $y\in E\oplus\Gamma E$ from \eqref{eq:relation of M_X and M'_X(1)}
and \eqref{eq:relation of M_X and M'_X(2)}. This implies that $M'_{X}$
satisfies \eqref{eq:implements_automorph(differential_ver.)}. Since
$M_{X}$ is uniquely determined except for constant numbers i.e. $M'_{X}=M_{X}+\mathrm{Constant}$
(see Theorem \ref{thm:implementability_condition}), we have\[
[M_{X},\pi_{P_{+}}(B(x))]=[M'_{X},\pi_{P_{+}}(B(x))],\quad x\in E\oplus\Gamma E.\]
On the other hand,\begin{align*}
[M_{X},B(x)] & =\{ M_{X}\pi_{P_{+}}(B(x))-\pi_{P_{+}}(B(x))M'_{X}\}+\pi_{P_{+}}(B(x))M'_{X}\\
 & \qquad-M'_{X}\pi_{P_{+}}(B(x))+\{ M'_{X}\pi_{P_{+}}(B(x))-\pi_{P_{+}}(B(x))M_{X}\}\\
 & =\pi_{P_{+}}(B(Xx))-[M'_{X},\pi_{P_{+}}(B(x))]+\pi_{P_{+}}(B(Xx)).\end{align*}
 Therefore we obtain \eqref{eq:implements_auto([M_X,B(f)]_type)}.
\end{proof}

\section{Proof of implementability}

Put\[
T(x,y)z:=\left(P\Gamma y,z\right)_{0}Px-\left(P\Gamma x,z\right)_{0}Py\]
for $x$, $y\in K$, $z\in E$. Then, for any $x$, $y\in K$, we
have\[
\pi_{P_{+}}(B(x)B(y))|_{\mathcal{E}_{+}}=\sum_{i,j=0}^{1}\Xi_{i,j}(\lambda_{i,j}),\]
where\begin{gather*}
\lambda_{0,0}=\left(P\Gamma x,Py\right)_{0},\\
\lambda_{1,0}=Px\wedge Py,\quad\lambda_{0,1}=PJ\Gamma x\wedge PJ\Gamma y,\\
\lambda_{1,1}=\left(1_{2}\otimes d\Gamma(T(x,y))^{(2)}\right)^{*}\tau.\end{gather*}
Put $M_{X}^{+}=\sum_{l,m=0}^{\infty}\Xi_{l,m}(\kappa_{l,m})$. Then
we have\begin{align}
[ & M_{X}^{+},\pi_{P_{+}}(B(x)B(y))]\nonumber \\
 & =\Xi_{0,0}(\kappa_{0,1}\circ_{1}\lambda_{1,0}-\lambda_{0,1}\circ_{1}\kappa_{1,0})\nonumber \\
 & +\sum_{l=1}^{\infty}\Xi_{l,0}\left(S_{0}^{l}\,_{0}^{0}(\kappa_{l,1}\circ_{1}\lambda_{1,0})-(l+1)S_{0}^{0}\,_{0}^{l}(\lambda_{0,1}\circ_{1}\kappa_{l+1,0})-lS_{0}^{1}\,_{0}^{l-1}(\lambda_{1,1}\circ_{1}\kappa_{l,0})\right)\nonumber \\
 & +\sum_{m=1}^{\infty}\Xi_{0,m}\left((m+1)S_{m}^{0}\,_{0}^{0}(\kappa_{0,m+1}\circ_{1}\lambda_{1,0})-S_{0}^{0}\,_{m}^{0}(\lambda_{0,1}\circ_{1}\kappa_{1,m})+mS_{m-1}^{0}\,_{1}^{0}(\kappa_{0,m}\circ_{1}\lambda_{1,1})\right)\nonumber \\
 & +\sum_{l,m=1}^{\infty}\Xi_{l,m}\left((m+1)S_{m}^{l}\,_{0}^{0}(\kappa_{l,m+1}\circ_{1}\lambda_{1,0})-(l+1)S_{0}^{0}\,_{m}^{l}(\lambda_{0,1}\circ_{1}\kappa_{l+1,m})\right.\nonumber \\
 & \left.+mS_{m}^{l}\,_{0}^{0}(\kappa_{l,m+1}\circ_{1}\lambda_{1,0})-lS_{0}^{1}\,_{m}^{l-1}(\lambda_{1,1}\circ_{1}\kappa_{l,m})\right).\label{eq:commutant_of_M_and_B(x)B(y)}\end{align}
by using \eqref{eq:composition_of_two_int_ker_ops}. On the other hand,
\begin{align}
\pi_{P_{+}} & (B(Xx)B(y)+B(x)B(Xy))\nonumber \\
 & =\Xi_{0,0}\left(\left\langle PJ\Gamma Xx,Py\right\rangle -\left\langle PJ\Gamma x,PJ\Gamma Xy\right\rangle \right)\nonumber \\
 & +\Xi_{1,0}\left(PXx\wedge Py+Px\wedge PXy\right)+\Xi_{0,1}\left(PXx\wedge Py+Px\wedge PXy\right)\nonumber \\
 & +\Xi_{1,1}\left(\left(1_{2}\otimes d\Gamma(T(Xx,y))^{(2)}+1_{2}\otimes d\Gamma(T(x,Xy))^{(2)}\right)^{*}\tau\right).\label{eq:pi_P(B(Xx)B(y)+B(x)B(Xy))}\end{align}
Thus \eqref{eq:implements_automorph(differential_ver)-No2} implies\begin{eqnarray}
 & \mathcal{A}_{l,0}\left(S_{0}^{l}\,_{0}^{0}(\kappa_{l,1}\circ_{1}\lambda_{1,0})-(l+1)S_{0}^{0}\,_{0}^{l}(\lambda_{0,1}\circ_{1}\kappa_{l+1,0})-lS_{0}^{1}\,_{0}^{l-1}(\lambda_{1,1}\circ_{1}\kappa_{l,0})\right)=0,\label{eq:coefficient_(l,0)}\\
 & \mathcal{A}_{0,m}\left((m+1)S_{m}^{0}\,_{0}^{0}(\kappa_{0,m+1}\circ_{1}\lambda_{1,0})-S_{0}^{0}\,_{m}^{0}(\lambda_{0,1}\circ_{1}\kappa_{1,m})+mS_{m-1}^{0}\,_{1}^{0}(\kappa_{0,m}\circ_{1}\lambda_{1,1})\right)=0\label{eq:coefficient_(0,m)}\end{eqnarray}
for $l$, $m\geq2$ and\begin{align}
\mathcal{A}_{l,m} & \left((m+1)S_{m}^{l}\,_{0}^{0}(\kappa_{l,m+1}\circ_{1}\lambda_{1,0})-(l+1)S_{0}^{0}\,_{m}^{l}(\lambda_{0,1}\circ_{1}\kappa_{l+1,m})\right.\nonumber \\
 & \left.+mS_{m}^{l}\,_{0}^{0}(\kappa_{l,m+1}\circ_{1}\lambda_{1,0})-lS_{0}^{1}\,_{m}^{l-1}(\lambda_{1,1}\circ_{1}\kappa_{l,m})\right)=0\label{eq:coefficient_(l,m)}\end{align}
for $l\geq1$, $m\geq1$, $(l,m)\neq(1,1)$. 

\begin{lem}
$\kappa_{l,0}=0$, $\kappa_{l,1}=0$, $\kappa_{0,m}=0$ and $\kappa_{1,m}=0$
for $l\geq2$ and $m\geq2$.
\end{lem}
\begin{proof}
\eqref{eq:coefficient_(l,0)} implies\[
0=\sum_{\mathbf{i}}\left\langle \kappa_{l,1},e(\mathbf{i})\otimes\lambda_{1,0}\right\rangle \mathcal{A}_{l,0}(e(\mathbf{i}))=\sum_{\mathbf{i}}\left\langle \kappa_{l,1},\mathcal{A}_{l,0}(e(\mathbf{i}))\otimes\lambda_{1,0}\right\rangle \mathcal{A}_{l,0}(e(\mathbf{i}))\]
for all $l\geq2$ since $\lambda_{1,0}=e_{\alpha}\wedge e_{\beta}$,
$\lambda_{0,1}=0$, and $\lambda_{1,1}=0$ for $x=e_{\alpha}$ and
$y=e_{\beta}$ $(\alpha<\beta)$. Thus we have $\left\langle \kappa_{l,1},\mathcal{A}_{l,0}(e(\mathbf{i}))\otimes\lambda_{1,0}\right\rangle =0$
for all $\mathbf{i}\in\mathbf{N}^{l}$, i.e., \begin{equation}
\kappa_{l,1}=0\label{eq:kappa_(l,1)_equal_0}\end{equation}
 for all $l\geq2$. On the other hand, \eqref{eq:coefficient_(l,0)}
implies \begin{equation}
\kappa_{l+1,0}=0\label{eq:kappa_(l+1,0)_equal_0}\end{equation}
 for all $l\geq2$ since $\lambda_{1,0}=0$, $\lambda_{0,1}=e_{\alpha}\wedge e_{\beta}$,
and $\lambda_{1,1}=0$ for $x=\Gamma e_{\alpha}$ and $y=\Gamma e_{\beta}$.
Thus we obtain\[
\kappa_{l,0}=0\]
for all $l\geq2$ by using \eqref{eq:coefficient_(l,0)}, \eqref{eq:kappa_(l,1)_equal_0},
and \eqref{eq:kappa_(l+1,0)_equal_0}. In the same manner, we have\[
\kappa_{0,m}=0,\quad\kappa_{1,m}=0\]
for $m\geq2$ from \eqref{eq:coefficient_(0,m)}. 
\end{proof}
\begin{lem}
\label{thm:kappa_(l,m)_equal_0}$\kappa_{l,m}=0$ for $(l,m)\in\mathbf{Z}_{\geq0}^{2}$
satisfying $l\geq2$ or $m\geq2$. 
\end{lem}
\begin{proof}
We prove this lemma by induction. We have already shown $\kappa_{1,m}=0$
for $m\geq2$. Assume $\kappa_{l,m}=0$ for $m\geq2$ and we prove
$\kappa_{l+1,m}=0$ for $m\geq2$. Since $\kappa_{l,m+1}=0$, \eqref{eq:coefficient_(l,m)}
implies\[
\sum_{\mathbf{i},\mathbf{j}}\left\langle \kappa_{l+1,m},\mathcal{A}_{2(l+1)}(\lambda_{0,1}\otimes e(\mathbf{i}))\otimes\mathcal{A}_{2m}(e(\mathbf{j}))\right\rangle \mathcal{A}_{2l}(e(\mathbf{i}))\otimes\mathcal{A}_{2m}(e(\mathbf{j}))=0,\]
i.e.,\[
\left\langle \kappa_{l+1,m},\mathcal{A}_{2(l+1)}((e_{\alpha}\otimes e_{\beta})\otimes e(\mathbf{i}))\otimes\mathcal{A}_{2m}(e(\mathbf{j}))\right\rangle =0\]
for all $\alpha$, $\beta$, $\mathbf{i}$, and $\mathbf{j}$. In
fact, $\lambda_{0,1}=e_{\alpha}\wedge e_{\beta}$ if $x=\Gamma e_{\alpha}$
and $y=\Gamma e_{\beta}$. Thus we have $\kappa_{l+1,m}=0$ for $m\geq2$.
By induction, $\kappa_{l,m}=0$ for $l\geq1$ and $m\geq2$. 
\end{proof}
Now we consider $\kappa_{1,1}$, $\kappa_{1,0}$, and $\kappa_{0,1}$.
We have\begin{align}
\sum_{\mathbf{i}}\langle\kappa_{1,1}, & e(\mathbf{i})\otimes(P_{+}x\wedge P_{+}y)\rangle e(\mathbf{i})-d\Gamma(T(x,y))^{(2)}\kappa_{1,0}\nonumber \\
 & =P_{+}Xx\wedge P_{+}y+P_{+}x\wedge P_{+}Xy\label{eq:coefficient_(1,0)}\end{align}
and\begin{align}
-\sum_{\mathbf{i}}\langle\kappa_{1,1}, & (P_{+}J\Gamma x\wedge P_{+}J\Gamma y)\otimes e(\mathbf{i})\rangle e(\mathbf{i})+\left(d\Gamma(T(x,y))^{(2)}\right)^{*}\kappa_{0,1}\nonumber \\
 & =P_{+}J\Gamma Xx\wedge P_{+}J\Gamma y+PJ\Gamma x\wedge PJ\Gamma Xy\label{eq:coefficient_(0,1)}\end{align}
for $x$, $y\in K$ by using \eqref{eq:implements_automorph(differential_ver)-No2},
\eqref{eq:commutant_of_M_and_B(x)B(y)} and \eqref{eq:pi_P(B(Xx)B(y)+B(x)B(Xy))}.

\begin{lem}
\label{thm:kappa_(1,1)_equal_etc}\begin{equation}
\kappa_{1,1}=\left(1_{2}\otimes d\Gamma(P_{+}XP_{+})^{(2)}\right)^{*}\tau.\label{eq:kappa_(1,1)_equal_etc}\end{equation}
$\kappa_{1,1}$ is in $(E^{\wedge2})\otimes(E^{\wedge2})^{*}$ if
$P_{+}XP_{+}$ is in $\mathcal{L}(E,E)$. Moreover, $\kappa_{1,1}$
is in $(H^{\wedge2})\otimes(E^{\wedge2})^{*}$ if $P_{+}XP_{+}$ is
Hilbert-Schmidt class operator.
\end{lem}
\begin{proof}
We have\[
\sum_{\mathbf{i}}\left\langle \kappa_{1,1},e(\mathbf{i})\otimes(x\wedge y)\right\rangle e(\mathbf{i})=d\Gamma(P_{+}XP_{+})^{(2)}(x\wedge y)\]
from \eqref{eq:coefficient_(1,0)} when $x\in P_{+}K$ and $y\in P_{+}K$.
Thus we obtain \eqref{eq:kappa_(1,1)_equal_etc}. 
\end{proof}
\eqref{eq:coefficient_(1,0)} implies\begin{equation}
d\Gamma(T(x,y))^{(2)}\kappa_{1,0}=-x\wedge P_{+}Xy\label{eq:T(x,y)kappa_(1,0)_equal_x_wedge_PXy}\end{equation}
for $x\in P_{+}K$ and $y\in P_{-}K$. Now put $x=e_{n}$ and $y=\Gamma e_{n}$.
Then \eqref{eq:T(x,y)kappa_(1,0)_equal_x_wedge_PXy} implies \[
\sum_{m;m\neq n}\left\langle \kappa_{1,0},e_{m}\wedge e_{n}\right\rangle e_{m}\wedge e_{n}=PX\Gamma e_{n}\wedge e_{n}.\]
Thus $\kappa_{1,0}$ and $\kappa_{0,1}$ is given by the formula\begin{gather}
\kappa_{1,0}=\frac{1}{2}\sum_{n\in\mathbf{N}}(P_{+}XP_{-}\Gamma e_{n})\wedge e_{n}\in\left(E^{\wedge2}\right)^{*},\label{eq:kappa_(1,0)_equal_etc}\\
\kappa_{0,1}=\frac{1}{2}\sum_{n\in\mathbf{N}}(P_{+}JXJP_{-}\Gamma e_{n})\wedge e_{n}\in H^{\wedge2}\label{eq:kappa_(0,1)_equal_etc}\end{gather}
 formally. Hence we can prove the following lemma.

\begin{lem}
\label{thm:kappa_(1,0)_and_kappa_(0,1)}$X\in o(K,\Gamma)$ (resp.
$X\in o(K,\Gamma;E)$) satisfies \eqref{eq:implemet_cond(H.S.cond)}
(resp. \eqref{eq:implement_cond(test_func_to_gen_func)}) if and only
if $\kappa_{1,0}\in\left(E^{\wedge2}\right)^{*}$ and $\kappa_{0,1}\in H^{\wedge2}$
(resp. $\kappa_{1,0}$, $\kappa_{0,1}\in\left(E^{\wedge2}\right)^{*}$).
\end{lem}
\begin{proof}
Suppose \eqref{eq:implemet_cond(H.S.cond)} (resp. \eqref{eq:implement_cond(test_func_to_gen_func)}).
Then, since\[
|\kappa_{l,m}|_{p}^{2}\leq\frac{1}{2}\sum_{n\in\mathbf{N}}|P_{+}XP_{-}\Gamma e_{n}|_{p}^{2}|e_{n}|_{p}^{2}\leq\frac{1}{2}\sum_{n\in\mathbf{N}}\lambda_{n}^{2p}|P_{+}XP_{-}\Gamma e_{n}|_{p}^{2}\]
for $p=0$ (resp. some $p>0$), we have $\kappa_{1,0}\in\left(E^{\wedge2}\right)^{*}$
and $\kappa_{0,1}\in H^{\wedge2}$(resp. $\kappa_{1,0}$, $\kappa_{0,1}\in\left(E^{\wedge2}\right)^{*}$).

Next, assume $\kappa_{1,0}\in\left(E^{\wedge2}\right)^{*}$ and $\kappa_{0,1}\in H^{\wedge2}$(resp.
$\kappa_{1,0}$, $\kappa_{0,1}\in\left(E^{\wedge2}\right)^{*}$).
Then\[
\sum_{n\in\mathbf{N}}P_{+}XP_{-}\Gamma e_{n}\otimes e_{n}\in H^{\otimes2}\]
follows from definition of $H^{\wedge2}$.(resp.\[
\sum_{n\in\mathbf{N}}P_{+}XP_{-}\Gamma e_{n}\otimes e_{n}\in\left(E^{\otimes2}\right)^{*}\]
follows from definition of $\left(E^{\wedge2}\right)^{*}$.) Therefore
\eqref{eq:implemet_cond(H.S.cond)} (resp. \eqref{eq:implement_cond(test_func_to_gen_func)})
holds. 
\end{proof}
From lemma \ref{thm:kappa_(l,m)_equal_0}, \ref{thm:kappa_(1,1)_equal_etc},
and \ref{thm:kappa_(1,0)_and_kappa_(0,1)}, we obtain lemma \ref{thm:implementability_ condition(odd_even_decomp_ver)}
for the case $\sigma=+$. Let us prove lemma \ref{thm:implementability_ condition(odd_even_decomp_ver)}
for the case $\sigma=-$. 

\begin{lem}
Let $f\in H=P_{+}K$ and $w(f):K\rightarrow K$ be a linear operator
defined by\[
w(f)x:=x-\left((1+\Gamma)f,x\right)_{K}(1+\Gamma)f,\quad x\in K.\]

\begin{enumerate}
\item $w(f)^{2}=1$ and $w(f)^{\dagger}=w(f)$, that is, $w(f)$ is a unitary
operator on $K$. Moreover $w(f)$ commutes with $\Gamma$. Therefore
$w(f)\in O(K,\Gamma)$.
\item Let $f\in E$ satisfy $(f,f)_{K}=1$. Then we have\[
W(f)q_{P_{+}}(Y)W(f)=q_{P_{+}}(Y(2w(f)-1))\]
for all $Y\in o_{\mathrm{fin}}(K,\Gamma)$. 
\end{enumerate}
\end{lem}
\begin{proof}
(1) follows from direct computation. 

(2) Note that\[
W(f):=a^{\dagger}(f)+a(Jf)=\pi_{P_{+}}(B((1+\Gamma)f))\]
follows $W(f)^{2}=1$ from Lemma \ref{thm:anti-commutation-relations}
and \[
[q(Y),B(x)]=B(Yf)\]
holds for all $x\in K$ and $Y\in o_{\mathrm{fin}}(K,\Gamma)$ on
$K$. (See Theorem 4.4 (6) of \cite{Araki_Contemp_math}.) From the
above note, we have\[
W(f)q_{P_{+}}(Y)W(f)=q_{P_{+}}(Y)-\pi_{P_{+}}(B(Y(1+\Gamma)f)B((1+\Gamma)f)).\]
Now, \[
q(Y(w(f)-1))=-\frac{1}{2}B(Y(1+\Gamma)f)B((1+\Gamma)f)\]
holds. This is the quantization of the following finite rank operator
:\[
Y(w(f)-1)x=-((1+\Gamma)f,x)_{K}Y(1+\Gamma)f,\quad x\in K.\]
Therefore\[
W(f)q_{P_{+}}(Y)W(f)=q_{P_{+}}(Y)+2q_{P_{+}}(Y(w(f)-1))=q_{P_{+}}(Y(2w(f)-1)).\]

\end{proof}
Let $M_{X}^{-}\in\mathcal{L}(\Gamma_{-}(H),\mathcal{E}_{-}^{*})$
satisfy \eqref{eq:implements_automorph(diff_ver)-odd_and_even_case}
on $\mathcal{E}_{-}$. Note that \[
W(f)M_{X}^{-}W(f)\in\mathcal{L}(\Gamma_{+}(H),\mathcal{E}_{+}^{*})\]
 and \[
W(f)q_{P_{+}}(Y)W(f)|_{\mathcal{E}_{+}}=q_{P_{+}}(Y(2w(f)-1))|_{\mathcal{E}_{+}}\in\mathcal{L}(\mathcal{E}_{+},\Gamma_{+}(H)).\]
Then the restriction of \eqref{eq:implements_automorph(differential_ver.)}
to $\mathcal{E}_{-}$ leads us to\begin{equation}
[W(f)M_{X}^{-}W(f),q_{P_{+}}(Y(2w(f)-1))]=q_{P_{+}}([X,Y](2w(f)-1))\label{eq:W(f)MW(f)_commute_q(A(1-C))_equal_q([X,A](1-C))}\end{equation}
for all $Y\in o_{\mathrm{fin}}(K,\Gamma)$ on $\mathcal{E}_{+}$.
Put\[
W(f)M_{X}^{-}W(f)=\sum_{l,m=0}^{\infty}\Xi_{l,m}(\kappa_{l,m}).\]
 \eqref{eq:W(f)MW(f)_commute_q(A(1-C))_equal_q([X,A](1-C))} implies\begin{align*}
\Bigg[\sum_{l,m=0}^{\infty} & \Xi_{l,m}(\kappa_{l,m}),\pi_{P_{+}}\Bigl(B(x)B((2w(f)-1)y)-B(y)B((2w(f)-1)x)\Bigl)\Bigg]\\
 & =\pi_{P_{+}}\Bigl(B(Xf)B((2w(f)-1)y)-B(Xy)B((2w(f)-1)y)\\
 & \quad+B(x)B((2w(f)-1)Xy)-B(y)B((2w(f)-1)Xx)\Bigl)\end{align*}
for all $x$, $y\in K$ on $\mathcal{E}_{+}$ when $Y\in o_{\mathrm{fin}}(K,\Gamma)$
is given by \eqref{eq:A_equal_H_1+sqrt(-1)H_2}. Therefore we can obtain\[
\kappa_{l,m}=0\]
for all $(l,m)$ satisfying $l\geq2$ or $m\geq2$. Moreover\begin{gather}
S_{0}^{1}\,_{0}^{0}(\kappa_{1,1}\circ_{1}\lambda'_{1,0})-S_{0}^{1}\,_{0}^{0}(\lambda'_{1,1}\circ_{1}\kappa_{1,0})=\lambda''_{1,0},\label{eq:(1,0)_coefficient_equal(odd_case)}\\
-S_{0}^{0}\,_{1}^{0}(\lambda'_{0,1}\circ_{1}\kappa_{1,1})+S_{0}^{0}\,_{1}^{0}(\kappa_{0,1}\circ_{1}\lambda'_{1,1})=\lambda''_{0,1},\label{eq:(0,1)_coefficient_equal(odd_case)}\\
S_{0}^{1}\,_{1}^{0}(\kappa_{1,1}\circ_{1}\lambda'_{1,1})-S_{0}^{1}\,_{1}^{0}(\lambda'_{1,1}\circ_{1}\kappa_{1,1})=\lambda''_{1,1},\label{eq:(1,1)_coefficient_equal(odd_case)}\end{gather}
where\begin{gather*}
2\lambda'_{1,0}:=P_{+}x\wedge P_{+}(2w(f)-1)y+P_{+}(2w(f)-1)x\wedge P_{+}y,\\
2\lambda'_{0,1}:=P_{+}J\Gamma x\wedge P_{+}J\Gamma(2w(f)-1)y+P_{+}J\Gamma(2w(f)-1)x\wedge P_{+}J\Gamma y,\\
2\lambda'_{1,1}:=\left(1_{2}\otimes d\Gamma[T(x,(2w(f)-1)y)+T((2w(f)-1)x,y)]^{(2)}\right)^{*}\tau,\\
2\lambda''_{1,0}:=P_{+}Xx\wedge P_{+}(2w(f)-1)y+P_{+}(2w(f)-1)Xx\wedge P_{+}y\\
+P_{+}x\wedge P_{+}(2w(f)-1)Xy+P_{+}(2w(f)-1)x\wedge P_{+}Xy,\\
2\lambda''_{0,1}:=P_{+}J\Gamma Xx\wedge P_{+}J\Gamma(2w(f)-1)y+P_{+}J\Gamma(2w(f)-1)Xx\wedge P_{+}J\Gamma y\\
+P_{+}J\Gamma x\wedge P_{+}J\Gamma(2w(f)-1)Xy+P_{+}J\Gamma(2w(f)-1)x\wedge P_{+}J\Gamma Xy,\\
2\lambda''_{1,1}:=(1_{2}\otimes[T(Xx,(2w(f)-1)y)+T((2w(f)-1))Xx,y)\\
+T(x,(2w(f)-1)Xy)+T((2w(f)-1)x,Xy)]^{(2)})^{*}\tau.\end{gather*}

\begin{lem}
\label{thm:property_of_(2w(f)-1)}We denote the orthogonal complement
of $\mathrm{span}[\{ f,\Gamma f\}]$ with respect to the Hilbert space
$(K,(\cdot,\cdot)_{K})$ by $\mathrm{span}[\{ f,\Gamma f\}]^{\perp}$.
\begin{enumerate}
\item $w(f)x=x$ for $x\in\mathrm{span}[\{ f,\Gamma f\}]^{\perp}$.
\item $Jx\in\mathrm{span}[\{ f,\Gamma f\}]^{\perp}$ for $x\in\mathrm{span}[\{ f,\Gamma f\}]^{\perp}$.
\item $T(x,y)=0$ and $T(\Gamma x,\Gamma y)=0$ for $x$, $y\in H=P_{+}K$. 
\end{enumerate}
\end{lem}
Lemma \ref{thm:property_of_(2w(f)-1)} is easily checked.

\begin{lem}
$\,$\begin{equation}
\kappa_{1,1}=\left(1_{2}\otimes d\Gamma(P_{+}w(f)Xw(f)P_{+})^{(2)}\right)^{*}\tau.\label{eq:kappa_(1,1)_equal_etc(odd_case)}\end{equation}

\end{lem}
\begin{proof}
Put $x:=\Gamma Jx'$, $y:=\Gamma Jy'$ for $x'$, $y'\in H\cap\mathrm{span}[\{ f,\Gamma f\}]^{\perp}$.
Then \eqref{eq:(1,0)_coefficient_equal(odd_case)} and Lemma \ref{thm:property_of_(2w(f)-1)}
imply\[
\sum_{\mathbf{i}}\left\langle \kappa_{1,1},e(\mathbf{i})\otimes(x'\wedge y')\right\rangle e(\mathbf{i})=d\Gamma(P_{+}w(f)Xw(f)P_{+})^{(2)}(x'\wedge y').\]
In the same manner, put $x:=\Gamma Jx'$ for $x'\in H\cap\mathrm{span}[\{ f,\Gamma f\}]^{\perp}$
and $y=f$, then we have\[
\sum_{\mathbf{i}}\left\langle \kappa_{1,1},e(\mathbf{i})\otimes(x'\wedge f)\right\rangle e(\mathbf{i})=d\Gamma(P_{+}w(f)Xw(f)P_{+})^{(2)}(x'\wedge f).\]
Thus $\kappa_{1,1}$ is given by \eqref{eq:kappa_(1,1)_equal_etc(odd_case)}.
\end{proof}
\begin{lem}
Let $X\in o(K,\Gamma)$ (resp. $X\in o(K,\Gamma;E)$) satisfy \eqref{eq:implemet_cond(H.S.cond)}
(resp. \eqref{eq:implement_cond(test_func_to_gen_func)}). Then\begin{gather}
\kappa_{1,0}=\frac{1}{2}P_{+}Xe_{1}\wedge e_{1}+\frac{1}{2}\sum_{n\in\mathbf{N}\backslash\{1\}}P_{+}w(e_{1})Xw(e_{1})P_{-}\Gamma e_{n}\wedge e_{n},\label{eq:odd_case_of_kappa_(1,0)}\\
\kappa_{0,1}=\frac{1}{2}P_{+}JXJe_{1}\wedge e_{1}+\frac{1}{2}\sum_{n\in\mathbf{N}\backslash\{1\}}P_{+}Jw(e_{1})Xw(e_{1})JP_{-}\Gamma e_{n}\wedge e_{n}\label{eq:odd_case_of_kappa_(0,1)}\end{gather}
and $\kappa_{1,0}\in\left(E^{\wedge2}\right)^{*}$ and $\kappa_{0,1}\in H^{\wedge2}$
(resp. $\kappa_{1,0}$, $\kappa_{0,1}\in\left(E^{\wedge2}\right)^{*}$). 
\end{lem}
\begin{proof}
If we put $x=e_{n}$, $y=\Gamma e_{n}$ ($n\neq1$), then\[
\lambda''_{1,0}=-P_{+}w(e_{1})X\Gamma e_{n}\wedge e_{n}\]
from Lemma \ref{thm:property_of_(2w(f)-1)}. On the other hand, if
$x=e_{1}$ and $y=\Gamma e_{1}$, then\begin{align*}
2\lambda''_{1,0} & =P_{+}Xe_{1}\wedge(-2e_{1})+0+e_{1}\wedge P_{+}(2w(e_{1})-1)X\Gamma e_{1}+(-e_{1})\wedge P_{+}X\Gamma e_{1}\\
 & =-2P_{+}Xe_{1}\wedge e_{1}+e_{1}\wedge P_{+}X\Gamma e_{1}-e_{1}\wedge P_{+}X\Gamma e_{1}\\
 & =-2P_{+}Xe_{1}\wedge e_{1}.\end{align*}
Thus \eqref{eq:(1,0)_coefficient_equal(odd_case)} implies\[
d\Gamma(T(e_{n},\Gamma e_{n}))^{(2)}\kappa_{1,0}=P_{+}w(e_{1})X\Gamma e_{n}\wedge e_{n}=P_{+}w(e_{1})Xw(e_{1})\Gamma e_{n}\wedge e_{n},\]
i.e.\[
\sum_{m;m\neq n}\left\langle \kappa_{1,0},e_{m}\wedge e_{n}\right\rangle e_{m}\wedge e_{n}=P_{+}w(e_{1})Xw(e_{1})\Gamma e_{n}\wedge e_{n}\]
for all $n\neq1$ and\[
\sum_{m;m\neq1}\left\langle \kappa_{1,0},e_{1}\wedge e_{m}\right\rangle e_{1}\wedge e_{m}=P_{+}w(e_{1})Xw(e_{1})\Gamma e_{1}\wedge e_{1}.\]
Therefore\begin{align*}
\kappa_{1,0} & =\sum_{m=1}^{\infty}\sum_{n;n>m}\left\langle \kappa_{1,0},e_{m}\wedge e_{n}\right\rangle e_{m}\wedge e_{n}\\
 & =\frac{1}{2}P_{+}Xe_{1}\wedge e_{1}+\frac{1}{2}\sum_{n\in\mathbf{N}\backslash\{1\}}P_{+}w(e_{1})Xw(e_{1})P_{-}\Gamma e_{n}\wedge e_{n}.\end{align*}
 In the same manner, we also have \eqref{eq:odd_case_of_kappa_(0,1)}.
\end{proof}

\section{One particle and positive energy Fock representation of gauge algebra}

In this section, we deal with the one particle and the Fock representation
of gauge algebra. We discuss possibility of constructing of the Fock
representations of gauge algebra via implementability of Bogoliubov
automorphisms at Lie algebraic level. First, let $\mathbf{T}:=\mathbf{R}/\mathbf{Z}\simeq[0,2\pi]$.
(Here we identify 0 with $2\pi$.) 

Let $K:=\mathbf{C}^{N}\otimes L^{2}(\mathbf{T}^{2n-1})\otimes\mathbf{C}^{2}$
and $\Gamma$ be the anti-linear isomorphism on $K$ satisfying\[
\Gamma\left(u\otimes\left(\begin{array}{c}
f_{1}\\
f_{2}\end{array}\right)\right):=\overline{u}\otimes\left(\begin{array}{cc}
1 & 0\\
0 & -1\end{array}\right)\left(\begin{array}{c}
\overline{f_{1}}\\
\overline{f_{2}}\end{array}\right),\]
for $u\in\mathbf{C}^{N}$ and $f_{i}\in L^{2}(\mathbf{T}^{2n-1})$.
Here $\overline{u}$ and $\overline{f_{i}}$ stand for the natural
complex conjugate of $u$ and $f_{i}$ respectively. Let\[
\mathfrak{h}:=1_{N}\otimes\left(\begin{array}{cc}
-\sqrt{-1}\ \frac{\mathrm{d}}{\mathrm{d}x}\  & m\\
m & \sqrt{-1}\ \frac{\mathrm{d}}{\mathrm{d}x}\ \end{array}\right)\]
for $n=1$ and\[
\mathfrak{h}:=\left(\begin{array}{cc}
-\sqrt{-1}\ \overrightarrow{\sigma}\cdot\nabla & m1_{N}\\
m1_{N} & \sqrt{-1}\ \overrightarrow{\sigma}\cdot\nabla\end{array}\right)\]
for $n\geq2$. Here $m\geq0$ is mass, and $\sigma_{1}$, $\sigma_{2}$,
$\ldots$, $\sigma_{2n-1}$ are self-adjoint $N\times N$ matrices
representing the Euclidean Clifford algebra in $\mathbf{R}^{2n-1}$
and we put\[
\overrightarrow{\sigma}:=(\sigma_{1},\sigma_{2},\ldots,\sigma_{2n-1}),\quad\nabla:=(\partial_{1},\partial_{2},\ldots,\partial_{2n-1}).\]
We call $\mathfrak{h}$ the Dirac Hamiltonian on $K$. For example,
in case of $n=2$, we take\begin{equation}
\sigma_{1}=\left(\begin{array}{cc}
0 & 1\\
1 & 0\end{array}\right),\quad\sigma_{2}=\left(\begin{array}{cc}
0 & -\sqrt{-1}\\
\sqrt{-1} & 0\end{array}\right),\quad\sigma_{3}=\left(\begin{array}{cc}
1 & 0\\
0 & -1\end{array}\right).\label{eq:Pauli_Matrices}\end{equation}
These $\sigma_{i}$ are called the Pauli matrices. The Pauli matrices
satisfy\[
\sigma_{i}^{2}=1,\quad\sigma_{i}\sigma_{j}=-\sigma_{j}\sigma_{i}\,(i\neq j).\]

\subsection{The one particle representation of gauge group and gauge algebra}

Let $U\in C^{\infty}(\mathbf{T}^{2n-1},SO(N))$ and\[
\left[\pi(U)\left(\begin{array}{c}
f_{1}\\
f_{2}\end{array}\right)\right](x):=(U(x)\otimes1_{2})\left(\begin{array}{c}
f_{1}(x)\\
f_{2}(x)\end{array}\right),\quad x\in\mathbf{T}^{2n-1}\]
for $f_{i}\in\mathbf{C}^{N}\otimes L^{2}(\mathbf{T}^{2n-1})$. Then
$\pi(U)$ is a unitary operator on $K$ and $\pi(U)$ satisfies\[
\Gamma\pi(U)=\pi(U)\Gamma,\]
that is, $\pi(U)$ is a Bogoliubov transformation on $K$. (In particular,
we deal with the case $N=2$ in the following subsections.)

Let $\varphi\in C^{\infty}(\mathbf{T}^{2n-1},\mathbf{R})$ and\[
U_{\lambda}(x):=\left(\begin{array}{cc}
\cos(\lambda\varphi(x)) & -\sin(\lambda\varphi(x))\\
\sin(\lambda\varphi(x)) & \cos(\lambda\varphi(x))\end{array}\right)\]
for $\lambda\in\mathbf{R}$. Then $U_{\lambda}$ is an element of
$C^{\infty}(\mathbf{T}^{2n-1},SO(2))$ and $\pi(U_{\lambda})$ is
a Bogoliubov transformation on $K$. 

The infinitesimal generator of the one parameter unitary group $\{ U_{\lambda}\}_{\lambda\in\mathbf{R}}$
is\[
X(x):=-\varphi(x)\sigma_{2}.\]
We define\[
\left[\pi(X)\left(\begin{array}{c}
f_{1}\\
f_{2}\end{array}\right)\right](x):=(X(x)\otimes1_{2})\left(\begin{array}{c}
f_{1}(x)\\
f_{2}(x)\end{array}\right),\]
for $f_{i}\in\mathbf{C}^{N}\otimes L^{2}(\mathbf{T}^{2n-1})$. Then
$\pi$ is a Lie algebra homomorphism from $C^{\infty}(\mathbf{T}^{2n-1},\  so(2))$
to $o(K,\Gamma)$. We call $\pi$ the one particle representation
of $C^{\infty}(\mathbf{T}^{2n-1},so(2))$.

The aim of this section is to give criterion of constructing the representation
of gauge algebra on the positive energy Fermion Fock spaces.

\subsection{The case of massive ($m>0$) and $n=1$}

First, we give the projection $P_{+}$ on $K$ in order to discuss
the possibility of constructing the positive energy Fock representation
of gauge algebra. Let\[
\phi_{\alpha}^{\sigma}(x):=\left(\begin{array}{c}
d_{\sigma\alpha}\\
\sigma d_{-\sigma\alpha}\end{array}\right)\exp(\sqrt{-1}\alpha x),\quad\sigma\in\{+,-\},\:\alpha\in\mathbf{Z},\]
where \[
d_{\alpha}:=\sqrt{\frac{1}{2}\left(1+\frac{\alpha}{\sqrt{E(\alpha)}}\right)},\quad E(\alpha):=\alpha^{2}+m^{2}.\]

\begin{lem}
$\phi_{\alpha}^{\sigma}$ is an eigenvector for $\mathfrak{h}$ with
an eigenvalue $\sigma\sqrt{E(\alpha)}$. 
\end{lem}
\begin{proof}
Let $\sigma=+$. Then\begin{align*}
-\sqrt{-1}\frac{\mathrm{d}}{\mathrm{d}x}d_{\alpha}e^{\sqrt{-1}\alpha x}+md_{-\alpha}e^{\sqrt{-1}\alpha x} & =\left(\alpha+m\sqrt{\frac{\sqrt{E(\alpha)}-\alpha}{\sqrt{E(\alpha)}+\alpha}}\right)d_{\alpha}e^{\sqrt{-1}\alpha x}\\
 & =\left(\alpha+\frac{m^{2}}{\sqrt{E(\alpha)}+\alpha}\right)d_{\alpha}e^{\sqrt{-1}\alpha x}\\
 & =\sqrt{E(\alpha)}d_{\alpha}e^{\sqrt{-1}\alpha x}.\end{align*}
In the same manner, we have\[
md_{\alpha}e^{\sqrt{-1}\alpha x}+\sqrt{-1}\frac{\mathrm{d}}{\mathrm{d}x}d_{-\alpha}e^{\sqrt{-1}\alpha x}=\sqrt{E(\alpha)}d_{-\alpha}e^{\sqrt{-1}\alpha x}.?\]
That is, \[
\left(\begin{array}{cc}
-\sqrt{-1}\ \frac{\mathrm{d}}{\mathrm{d}x}\  & m\\
m & \sqrt{-1}\ \frac{\mathrm{d}}{\mathrm{d}x}\ \end{array}\right)\phi_{\alpha}^{+}=\sqrt{E(\alpha)}\phi_{\alpha}^{+}.\]
The case of $\sigma=-$ can be considered similarly.
\end{proof}
Now let $\{ e_{s}\}_{s=1}^{N}$ be a C.O.N.S. of $\mathbf{C}^{N}$.
Then\[
\{ e_{s}\otimes\phi_{\alpha}^{\sigma}\ |\ \alpha\in\mathbf{Z},\  s\in\{1,2,\ldots,N\},\ \sigma\in\{+,-\}\}\]
is a C.O.N.S. of $K:=\mathbf{C}^{N}\otimes L^{2}(\mathbf{T})\otimes\mathbf{C}^{2}$. 

Let $P_{+}$(resp. $P_{-}$) be a spectral projection of $\mathfrak{h}$
for positive spectrum $[m,\infty)$ (resp. negative spectrum $(-\infty,-m]$).
Then\[
\{ e_{s}\otimes\phi_{\alpha}^{+}\ |\ \alpha\in\mathbf{Z},\  s\in\{1,2,\ldots,N\}\}\]
is a C.O.N.S. of $P_{+}K$. Let $N=2$ and put a C.O.N.S. of $\mathbf{C}^{N}=\mathbf{C}^{2}$
as follows:\[
e_{+}:=\frac{1}{\sqrt{2}}\left(\begin{array}{c}
\sqrt{-1}\\
1\end{array}\right),\quad e_{-}:=\frac{1}{\sqrt{2}}\left(\begin{array}{c}
-\sqrt{-1}\\
1\end{array}\right).\]
 Then the C.O.N.S. $\{ e_{s}\otimes\phi_{\alpha}^{\sigma}\ |\ \alpha\in\mathbf{Z},\  s,\sigma\in\{+,-\}\}$
satisfies\[
\Gamma(e_{s}\otimes\phi_{\alpha}^{+})=e_{-s}\otimes\phi_{-\alpha}^{-}.\]
This shows $P_{+}\Gamma=\Gamma P_{-}$. 

We check the criterion of implementability by using the result of
section 3. In order to use the white noise calculus, we need a Hilbert
space $H$ and a self adjoint operator $A$ on $H$ satisfying the
two conditions (i) and (ii) of Definition \ref{thm:property_of_self-adj_op_A}.
Let $H:=P_{+}K$. If $m>1$, then $A:=\mathfrak{h}P_{+}$. If $0<m\leq1$,
then we put $A:=(\mathfrak{h}+cP_{+}-cP_{-})P_{+}$ for some $c>1$.
To simplify discussion, we assume $m>1$. 

A complex structure of $H$ is given by the following anti-linear
isomorphism $J$ on $K$ : \[
J\left(u\otimes\left(\begin{array}{c}
f_{1}\\
f_{2}\end{array}\right)\right):=\overline{u}\otimes\left(\begin{array}{cc}
0 & 1\\
1 & 0\end{array}\right)\left(\begin{array}{c}
\overline{f_{1}}\\
\overline{f_{2}}\end{array}\right),\]
for $u\in\mathbf{C}^{N}$ and $f_{i}\in L^{2}(\mathbf{T}^{2n-1})$.
Here $\overline{u}$ and $\overline{f_{i}}$ stand for the natural
complex conjugate of $u$ and $f_{i}$ respectively. In fact, $J$
commutes with $P_{+}$. Let $E$ be the CH-space determined by the
pair $(H,A)$. We have the following relation between the canonical
bilinear form $\left\langle \cdot,\cdot\right\rangle $ of $E^{*}\times E$
and the inner product of $H$:\[
\left\langle f,g\right\rangle =(Jf,g)_{0},\quad f\in H,\  g\in E.\]

The following lemma is easily checked by direct computation.

\begin{lem}
\label{thm:<e_n,PX(1-P)e_n>(n=3D1_massive)}$\,$\begin{gather*}
(e_{+}\otimes\phi_{\alpha}^{+},\  P_{+}\pi(X)P_{-}\Gamma(e_{-}\otimes\phi_{\beta}^{+}))_{K}=-\sqrt{-1}\ (d_{\alpha}d_{\beta}-d_{-\alpha}d_{-\beta})\widehat{\varphi}(\alpha+\beta),\\
(e_{+}\otimes\phi_{\alpha}^{+},\  P_{+}\pi(X)P_{-}\Gamma(e_{+}\otimes\phi_{\beta}^{+}))_{K}=(e_{-}\otimes\phi_{\alpha}^{+},\  P_{+}\pi(X)P_{-}\Gamma(e_{-}\otimes\phi_{\beta}^{+}))_{K}=0\end{gather*}
where $\widehat{\varphi}$ is the Fourier co-efficient of $\varphi$.
\end{lem}
Therefore, we have

\begin{prop}
$\,$
\begin{enumerate}
\item $\mathrm{ad}(X)$ is implementable as $\mathcal{L}(\Gamma(H),\mathcal{E}^{*})$,
that is, $P_{+}\pi(X)P_{-}$ is a Hilbert-Schmidt class operator.
\item For any $p\geq\frac{3}{4}$, we have\[
\sum_{\alpha\in\mathbf{Z},s\in\{+,-\}}E(\alpha)^{p}|P_{+}\pi(X)P_{-}\Gamma(e_{s}\otimes\phi_{\alpha}^{+})|_{p}^{2}=+\infty,\]
that is, $\mathrm{ad}(X)$ is not implementable as $\mathcal{L}(\mathcal{E},\mathcal{E})$. 
\end{enumerate}
\end{prop}
\begin{proof}
(1) From lemma \ref{thm:<e_n,PX(1-P)e_n>(n=3D1_massive)}, we have\begin{align}
\sum_{\alpha\in\mathbf{Z},\  s\in\{+,-\}} & E(\alpha)^{p}|P_{+}\pi(X)P_{-}\Gamma(e_{s}\otimes\phi_{\alpha}^{+})|_{p}^{2}\nonumber \\
 & =\sum_{\alpha\in\mathbf{Z},\  s\in\{+,-\}}E(\alpha)^{p}\sum_{\beta\in\mathbf{Z},\  t\in\{+,-\}}|(e_{t}\otimes\phi_{\beta}^{+},\  P_{+}\pi(X)P_{-}\Gamma(e_{s}\otimes\phi_{\alpha}^{+}))_{K}|^{2}\notag\\
 & =2\sum_{\gamma\in\mathbf{Z}}|\widehat{\varphi}(\gamma)|^{2}\sum_{\alpha\in\mathbf{Z}}E(\alpha)^{p}E(\gamma-\alpha)^{p}\nonumber \\
 & \qquad\left(1-\frac{\alpha(\alpha-\gamma)}{\sqrt{E(\alpha)E(\gamma-\alpha)}}-\frac{m^{2}}{\sqrt{E(\alpha)E(\gamma-\alpha)}}\right)\label{eq:p-H.S.norm_of_PX(1-P)}\end{align}
for any $p\in\mathbf{R}$. We prove that the right hand side of \eqref{eq:p-H.S.norm_of_PX(1-P)}
is finite for any $0\leq p<\frac{3}{4}$. Put\[
N_{\alpha}:=\sqrt{\left\{ 1+\left(\frac{m}{\alpha}\right)^{2}\right\} \left\{ \left(1-\frac{\gamma}{\alpha}\right)^{2}+\left(\frac{m}{\alpha}\right)^{2}\right\} }+\left(1-\frac{\gamma}{\alpha}\right)+\left(\frac{m}{\alpha}\right)^{2}.\]
Then

\begin{align*}
\sqrt{E(\alpha)E(\gamma-\alpha)}-\alpha(\alpha-\gamma)-m^{2} & =\frac{E(\alpha)E(\gamma-\alpha)-(\alpha(\alpha-\gamma)+m^{2})^{2}}{\sqrt{E(\alpha)E(\gamma-\alpha)}+\alpha(\alpha-\gamma)+m^{2}}\\
 & =N_{\alpha}^{-1}m^{2}\gamma^{2}\alpha^{-2}.\end{align*}
If $|\alpha|$ is larger than $\alpha_{0}:=\max\{|\gamma|+1,m\}$,
then $N_{\alpha}$ satisfies \[
\frac{2}{|\gamma|+1}\leq N_{\alpha}<7\]
In fact, \[
N_{\alpha}\geq\sqrt{(1+0^{2})\left\{ \left(1-\frac{|\gamma|}{|\gamma|+1}\right)^{2}+0^{2}\right\} }+\left(1-\frac{|\gamma|}{|\gamma|+1}\right)+0^{2}\geq\frac{2}{|\gamma|+1}\]
and\[
N_{\alpha}\leq\sqrt{(1+1)\{(1+1)^{2}+1\}}+(1+1)+1<7\]
for $\alpha\in\mathbf{Z}$ satisfying $|\alpha|\geq\alpha_{0}$. Then
we have\begin{align*}
\sum_{\alpha\in\mathbf{Z};|\alpha|\geq\alpha_{0}} & E(\alpha)^{p-\frac{1}{2}}E(\gamma-\alpha)^{p-\frac{1}{2}}(\sqrt{E(\alpha)E(\gamma-\alpha)}-\alpha(\alpha-\gamma)-m^{2})\\
 & \leq m^{2}|\gamma|^{2}\frac{|\gamma|+1}{2}\sum_{\alpha\in\mathbf{Z};|\alpha|\geq\alpha_{0}}E(\alpha)^{p-\frac{1}{2}}E(\gamma-\alpha)^{p-\frac{1}{2}}\alpha^{-2}\\
 & =m^{2}|\gamma|^{2}\frac{|\gamma|+1}{2}\sum_{\alpha\in\mathbf{Z};|\alpha|\geq\alpha_{0}}\alpha^{4p-4}\left(1+\left(\frac{m}{\alpha}\right)^{2}\right)^{p-\frac{1}{2}}\left(\left(1-\frac{\gamma}{\alpha}\right)^{2}+\left(\frac{m}{\alpha}\right)^{2}\right)^{p-\frac{1}{2}}\\
 & \leq2^{2p-2}m^{2}|\gamma|^{2}(|\gamma|+1)\sum_{\alpha\in\mathbf{Z}\backslash\{0\}}\alpha^{4p-4}.\end{align*}
On the other hand, \begin{align*}
\sum_{\alpha\in\mathbf{Z};|\alpha|<\alpha_{0}} & E(\alpha)^{p}E(\gamma-\alpha)^{p}\left(1-\frac{\alpha(\alpha-\gamma)}{\sqrt{E(\alpha)E(\gamma-\alpha)}}-\frac{m^{2}}{\sqrt{E(\alpha)E(\gamma-\alpha)}}\right)\\
 & \leq\sum_{\alpha\in\mathbf{Z};|\alpha|<\alpha_{0}}E(\alpha)E(\gamma-\alpha)(1+1+0)\\
 & \leq C_{2}|\gamma|^{2}+C_{1}|\gamma|+C_{0},\end{align*}
where all $C_{i}$ are polynomials of $\alpha_{0}=\max\{|\gamma|+1,m\}$,
that is, $C_{2}|\gamma|^{2}+C_{1}|\gamma|+C_{0}$ is a polynomial
of $|\gamma|$. 

Therefore, the right hand side of the following inequality\begin{align*}
\sum_{\alpha\in\mathbf{Z},s\in\{+,-\}} & E(\alpha)^{p}|P_{+}\pi(X)P_{-}\Gamma(e_{s}\otimes\phi_{\alpha}^{+})|_{p}^{2}\\
 & \leq\sum_{\gamma\in\mathbf{Z}}|\widehat{\varphi}(\gamma)|^{2}\left(C_{2}|\gamma|^{2}+C_{1}|\gamma|+C_{0}+2^{2p-2}m^{2}\gamma^{2}(|\gamma|+1)\sum_{\alpha\in\mathbf{Z}\backslash\{0\}}\alpha^{4p-4}\right)\end{align*}
is finite when $0\leq p<\frac{3}{4}$. 

If $p\geq\frac{3}{4}$, the right hand side of \eqref{eq:p-H.S.norm_of_PX(1-P)}
does not converge. Because, we have\begin{align*}
\sum_{\alpha\in\mathbf{Z};|\alpha|\geq\alpha_{0}} & E(\alpha)^{p-\frac{1}{2}}E(\gamma-\alpha)^{p-\frac{1}{2}}(\sqrt{E(\alpha)E(\gamma-\alpha)}-\alpha(\alpha-\gamma)-m^{2})\\
 & \geq\frac{1}{7}m^{2}|\gamma|^{2}\sum_{\alpha\in\mathbf{Z};|\alpha|\geq\alpha_{0}}\alpha^{4p-4}\left(1+\left(\frac{m}{\alpha}\right)^{2}\right)^{p-\frac{1}{2}}\left(\left(1-\frac{\gamma}{\alpha}\right)^{2}+\left(\frac{m}{\alpha}\right)^{2}\right)^{p-\frac{1}{2}}\\
 & \geq\frac{1}{7}m^{2}|\gamma|^{2}\sum_{\alpha\in\mathbf{Z};|\alpha|\geq\alpha_{0}}\alpha^{4p-4}\cdot(1+0)^{p-\frac{1}{2}}\cdot\left(\left(1-\frac{|\gamma|}{|\gamma|+1}\right)^{2}+0\right)^{p-\frac{1}{2}}\\
 & =\frac{1}{7}m^{2}|\gamma|^{2}\left(\frac{1}{|\gamma|+1}\right)^{2p-1}\sum_{\alpha\in\mathbf{Z};|\alpha|\geq\alpha_{0}}\alpha^{4p-4}\end{align*}
and $\sum_{\alpha\in\mathbf{Z};|\alpha|\geq\alpha_{0}}\alpha^{4p-4}$
does not converge for $p \geq \frac{3}{4}$. 
\end{proof}

\subsection{The case of massless ($m=0$) and $n=1$}

Let\[
\phi_{\alpha}^{+}(x):=\left(\begin{array}{c}
1\\
0\end{array}\right)\exp(\sqrt{-1}\alpha x),\quad\phi_{\alpha}^{-}(x):=\left(\begin{array}{c}
0\\
-1\end{array}\right)\exp(\sqrt{-1}\alpha x),\quad\alpha\in\mathbf{Z}\backslash\{0\}\]
and\[
\phi_{0}^{+}(x):=\frac{1}{\sqrt{2}}\left(\begin{array}{c}
1\\
1\end{array}\right),\quad\phi_{0}^{-}(x):=\frac{1}{\sqrt{2}}\left(\begin{array}{c}
1\\
-1\end{array}\right).\]
Then $\phi_{\sigma n}^{\sigma}$ is an eigenvector for the Dirac Hamiltonian
$\mathfrak{h}$ with an eigenvalue $n.$ Let \[
e_{+}:=\frac{1}{\sqrt{2}}\left(\begin{array}{c}
\sqrt{-1}\\
1\end{array}\right),\quad e_{-}:=\frac{1}{\sqrt{2}}\left(\begin{array}{c}
-\sqrt{-1}\\
1\end{array}\right).\]
$\{ e_{s}\}_{s=+,-}$ is a C.O.N.S. of $\mathbf{C}^{N}=\mathbf{C}^{2}.$
Then\[
\{ e_{s}\otimes\phi_{\alpha}^{\sigma}\ |\ \alpha\in\mathbf{Z},\ \sigma,\  s\in\{+,-\}\}\]
is a C.O.N.S. of $K:=\mathbf{C}^{N}\otimes L^{2}(\mathbf{T})\otimes\mathbf{C}^{2}$,
$N=2$.

Let\begin{gather*}
P_{+}K:=\overline{\mathrm{span}}\{ e_{s}\otimes\phi_{\alpha}^{\sigma},\ \  e_{s}\otimes\phi_{0}^{+}|\ \alpha\in\mathbf{N},\ \sigma,\  s\in\{+,-\}\},\\
P_{-}K:=\overline{\mathrm{span}}\{ e_{s}\otimes\phi_{\alpha}^{\sigma},\  e_{s}\otimes\phi_{0}^{-}|\ \alpha\in\mathbf{(-N)},\ \sigma,\  s\in\{+,-\}\}\end{gather*}
and $P_{+}$ (resp. $P_{-}$) be a projection from $K$ to $P_{+}K$
(resp. $P_{-}K$). Then\[
\Gamma(e_{s}\otimes\phi_{\alpha}^{\sigma})=e_{-s}\otimes\phi_{-\alpha}^{\sigma},\quad\Gamma(e_{s}\otimes\phi_{0}^{+})=e_{-s}\otimes\phi_{0}^{-},\]
for $\alpha\in\mathbf{N}$. This implies $P_{+}\Gamma=\Gamma P_{-}$. 

Let $H:=P_{+}K$ and $A:=(\mathfrak{h}+cP_{+}-cP_{-})P_{+}$ for some
$c>1$. Then we can define the CH-space $E$ by using the pair $(H,A)$. 

\begin{prop}
$\,$
\begin{enumerate}
\item $\mathrm{ad}(X)$ is implementable as $\mathcal{L}(\Gamma(H),\mathcal{E}^{*})$.
\item $\mathrm{ad}(X)$ is implementable as $\mathcal{L}(\mathcal{E},\mathcal{E})$. 
\end{enumerate}
\end{prop}
\begin{proof}
(1) Since $X(x)e_{s}=s\varphi(x)e_{s}$ for $s\in\{+,-\}$, we have\begin{align*}
(e_{s}\otimes\phi_{\alpha}^{\sigma},\ \pi(X)(e_{t}\otimes\phi_{\beta}^{\sigma'}))_{K} & =\int_{0}^{2\pi}\varphi(x)(e_{s},e_{t})_{\mathbf{C}^{2}}(\phi_{\alpha}^{\sigma}(x),\phi_{\beta}^{\sigma'}(x))_{\mathbf{C}^{2}}\frac{\mathrm{d}x}{2\pi}\\
 & =\left\{ \begin{array}{cc}
\delta_{s,t}\delta_{\sigma,\sigma'}\widehat{\varphi}(\alpha-\beta), & \beta\in\mathbf{Z}\backslash\{0\},\\
\frac{1}{\sqrt{2}}\delta_{s,t}\sigma\sigma'\widehat{\varphi}(\alpha), & \beta=0,\end{array}\right.\end{align*}
for any $\alpha\in\mathbf{Z}\backslash\{0\}$ and $\sigma$, $\sigma'$,
$s$, $t\in\{+,-\}$. From direct computation, \begin{align*}
|P_{+}\pi(X)P_{-}\Gamma|_{\mathrm{H.S.}}^{2} & =2|\widehat{\varphi}(0)|^{2}+4\sum_{\alpha\in\mathbf{N}\cup\{0\}}\sum_{\beta\in\mathbf{N}}|\widehat{\varphi}(\alpha+\beta)|^{2}\\
 & =2|\widehat{\varphi}(0)|^{2}+4\sum_{\gamma\in\mathbf{N}}\sum_{\alpha=0}^{\gamma-1}|\widehat{\varphi}(\gamma)|^{2}\\
 & \leq2|\widehat{\varphi}(0)|^{2}+\frac{1}{4}\sum_{\gamma\in\mathbf{N}}\sum_{\alpha=0}^{\gamma-1}\frac{1}{\gamma^{4}}\left\Vert \varphi^{(2)}\right\Vert _{L^{2}(\mathbf{T})}^{2}\\
 & =2|\widehat{\varphi}(0)|^{2}+\frac{1}{4}\left(\sum_{\gamma\in\mathbf{N}}\frac{1}{\gamma^{3}}\right)\left\Vert \varphi^{(2)}\right\Vert _{L^{2}(\mathbf{T})}^{2}<+\infty\end{align*}
where $\varphi^{(i)}$ is the $i$-th derivative of a function $\varphi$.
Here we use\[
\int_{0}^{2\pi}\varphi(x)\exp(-\sqrt{-1}\gamma x)\mathrm{d}x=\frac{1}{\sqrt{-1}\gamma}\int_{0}^{2\pi}\varphi^{(1)}(x)\exp(-\sqrt{-1}\gamma x)\mathrm{d}x.\]

(2) Recall\[
M_{X}=\Xi_{1,1}(\kappa_{1,1})+\Xi_{1,0}(\kappa_{1,0})+\Xi_{0,1}(\kappa_{0,1})+\mathrm{Constant}\]
 and $\kappa_{1,1}$ (resp. $\kappa_{1,0}$) is given by \eqref{eq:kappa_(1,1)_equal_etc}
(resp. \eqref{eq:kappa_(1,0)_equal_etc}). Thus we have to check $\kappa_{1,1}\in(E^{\wedge2})\otimes(E^{\wedge2})^{*}$
and $\kappa_{1,0}\in E^{\wedge2}$. 

First, we prove $\kappa_{1,1}\in(E^{\wedge2})\otimes(E^{\wedge2})^{*}$.
If $P_{+}\pi(X)P_{+}\in\mathcal{L}(E,E)$, then $\kappa_{1,1}\in(E^{\wedge2})\otimes(E^{\wedge2})^{*}$.
Thus it suffices to show $P_{+}\pi(X)P_{+}\in\mathcal{L}(E,E)$. For
$p\in\mathbf{N}$, since\[
A^{p}=P_{+}(\mathfrak{h}+cP_{+}-cP_{-})^{p}=P_{+}\sum_{q=0}^{p}\left(\begin{array}{c}
p\\
q\end{array}\right)c^{p-q}\mathfrak{h}^{q},\]
we have\begin{align*}
A^{p}\pi(X)\left(\begin{array}{c}
f_{1}\\
f_{2}\end{array}\right)=\sum_{q=0}^{p} & c^{p-q}\left(\begin{array}{c}
p\\
q\end{array}\right)\sum_{r=0}^{q}\left(\begin{array}{c}
q\\
r\end{array}\right)\\
 & (\sqrt{-1})^{q-r}\left(\begin{array}{cc}
(-1)^{q-r}X^{(q-r)} & 0\\
0 & X^{(q-r)}\end{array}\right)\mathfrak{h}^{r}\left(\begin{array}{c}
f_{1}\\
f_{2}\end{array}\right)\end{align*}
for $f_{i}\in C^{\infty}(\mathbf{T})$. Thus we obtain the following
estimation : \begin{align*}
\biggl|P_{+} & \pi(X)P_{+}\left(\begin{array}{c}
f_{1}\\
f_{2}\end{array}\right)\biggr|_{p}\\
 & \leq\sum_{q=0}^{p}c^{p-q}\left(\begin{array}{c}
p\\
q\end{array}\right)\sum_{r=0}^{q}\left(\begin{array}{c}
q\\
r\end{array}\right)\left|\left(\begin{array}{cc}
(-1)^{q-r}X^{(q-r)} & 0\\
0 & X^{(q-r)}\end{array}\right)\mathfrak{h}^{r}\left(\begin{array}{c}
f_{1}\\
f_{2}\end{array}\right)\right|_{0}\\
 & \leq C\left|\mathfrak{h}^{r}\left(\begin{array}{c}
f_{1}\\
f_{2}\end{array}\right)\right|_{0}\leq C\left|A^{r}\left(\begin{array}{c}
f_{1}\\
f_{2}\end{array}\right)\right|_{0}=C\left|\left(\begin{array}{c}
f_{1}\\
f_{2}\end{array}\right)\right|_{p},\end{align*}
where $X^{(q-r)}$ is the $(q-r)$-th derivative of a matrix valued
function $X$ and\[
C:=\sup_{0\leq q\leq p}\sup_{x\in\mathbf{T}}|X^{(q)}(x)|\cdot\sum_{q=0}^{p}c^{p-q}\left(\begin{array}{c}
p\\
q\end{array}\right)\sum_{r=0}^{q}\left(\begin{array}{c}
q\\
r\end{array}\right).\]
This estimation shows continuity of a linear operator $P_{+}\pi(X)P_{+}$
with respect to the topology of $E$.

Next, we prove $\kappa_{1,0}\in E^{\wedge2}$. \begin{align*}
|\kappa_{1,0}|_{p} & \leq4\sum_{\alpha\in\mathbf{N}}\sum_{\beta\in\mathbf{N}\cup\{0\}}(\alpha+c)^{p}(\beta+c)^{p}|\widehat{\varphi}(\alpha+\beta)|^{2}\\
 & =4\sum_{\gamma\in\mathbf{N}}\sum_{\alpha=0}^{\gamma-1}(\alpha+c)^{p}((\gamma-\alpha)+c)^{p}|\widehat{\varphi}(\gamma)|^{2}\\
 & \leq4\sum_{\gamma\in\mathbf{N}}\sum_{\alpha=0}^{\gamma-1}(\gamma+c)^{p}(\gamma+c)^{p}|\widehat{\varphi}(\gamma)|^{2}\\
 & \leq4\sum_{\gamma\in\mathbf{N}}(\gamma+c)^{2p+1}|\widehat{\varphi}(\gamma)|^{2}.\end{align*}
Now \[
\widehat{\varphi}(\gamma)=\int_{0}^{2\pi}\varphi(x)\exp(\sqrt{-1}\gamma x)\frac{\mathrm{d}x}{2\pi}=\left(\frac{-1}{\sqrt{-1}\gamma}\right)^{q}\int_{0}^{2\pi}\varphi^{(q)}(x)\exp(\sqrt{-1}\gamma x)\frac{\mathrm{d}x}{2\pi}\]
for any $q\in\mathbf{N}.$ This implies that $|\kappa_{1,0}|_{p}$
is finite. In fact, it can be easily seen that \[
\sum_{\gamma\in\mathbf{N}}(\gamma+c)^{2p+1}|\widehat{\varphi}(\gamma)|^{2}\]
 is finite for a sufficient large $q\in\mathbf{N}$. 
\end{proof}

\subsection{The case of massive ($m>0$) and $n=2$}

Let\begin{gather*}
\phi_{\alpha,s}^{+}(x)=\left(\begin{array}{c}
A_{\alpha}e_{s}\\
A_{-\alpha}e_{s}\end{array}\right)\exp(\sqrt{-1}(\alpha,x)),\\
\phi_{\alpha,s}^{-}(x)=\left(\begin{array}{c}
A_{-\alpha}e_{s}\\
-A_{\alpha}e_{s}\end{array}\right)\exp(\sqrt{-1}(\alpha,x)),\end{gather*}
where $x\in\mathbf{T}^{3}$, $\alpha\in\mathbf{Z}^{3}$, $s\in\{+,-\}$
and $\{ e_{+},e_{-}\}$ is a C.O.N.S. of $\mathbf{C}^{2}$. Moreover,
we put \[
A_{\alpha}:=\sqrt{\frac{1}{2}\left(1+\sum_{j=1}^{3}\frac{\alpha_{j}}{\sqrt{E(\alpha)}}\sigma_{j}\right)}\in\mathrm{Mat}(2,\mathbf{C}),\]
 where $\sigma_{1}$, $\sigma_{2}$, $\sigma_{3}$ are the Pauli matrices
and $E(\alpha):=|\alpha|^{2}+m^{2}$. (See \eqref{eq:Pauli_Matrices}.)
Since $1+\sum_{j=1}^{3}\frac{\alpha_{j}}{\sqrt{E(\alpha)}}\sigma_{j}$
is positive, we can define the square root of $1+\sum_{j=1}^{3}\frac{\alpha_{j}}{\sqrt{E(\alpha)}}\sigma_{j}$,
that is, $A_{n}$ is well-defined. 

\begin{lem}
\label{lem:A=3Dd_(+)p_(+) + d_(-)p_(-)}Let\[
d_{\alpha}^{\sigma}:=\sqrt{\frac{1}{2}\left(1+\sigma\frac{|\alpha|}{\sqrt{E(\alpha)}}\right)},\quad p_{\alpha}:=\frac{1}{2}\left(1+\sum_{j=1}^{3}\frac{\alpha_{j}}{|\alpha|}\sigma_{j}\right)\]
for $\alpha\in\mathbf{Z}^{3}\backslash\{0\}$ and $\sigma\in\{+,-\}$.
Then
\begin{enumerate}
\item $p_{\alpha}$ is a projection and $1-p_{\alpha}=p_{-\alpha}$.
\item $A_{\alpha}=d_{\alpha}^{+}p_{\alpha}+d_{\alpha}^{-}(1-p_{\alpha})$.
\end{enumerate}
\end{lem}
\begin{proof}
(1) follows from direct computation. 

(2) From direct computation, we have\[
A_{\alpha}^{2}=(d_{\alpha}^{+})^{2}p_{\alpha}+(d_{\alpha}^{-})^{2}(1-p_{\alpha}).\]
Thus \[
A_{\alpha}=\sqrt{(d_{\alpha}^{+})^{2}p_{\alpha}+(d_{\alpha}^{-})^{2}(1-p_{\alpha})}=d_{\alpha}^{+}p_{\alpha}+d_{\alpha}^{-}(1-p_{\alpha}).\]
 
\end{proof}
\begin{lem}
$\phi_{\alpha,s}^{\sigma}$ is an eigenvector for the Dirac operator
$\mathfrak{h}$ with an eigenvalue $\sigma\sqrt{E(\alpha)}$.
\end{lem}
\begin{proof}
From\[
\left(\sum_{j=1}^{3}\alpha_{j}\sigma_{j}\right)p_{\sigma\alpha}=\sigma|\alpha|p_{\sigma\alpha}\]
and Lemma \ref{lem:A=3Dd_(+)p_(+) + d_(-)p_(-)}, we have\[
\left(\sum_{j=1}^{3}\alpha_{j}\sigma_{j}\right)A_{\alpha}=|\alpha|(d_{\alpha}^{+}p_{\alpha}-d_{\alpha}^{-}p_{-\alpha}).\]
Thus\begin{align*}
-\sqrt{-1} & \sum_{j=1}^{3}\sigma_{j}\partial_{j}(A_{\alpha}e_{s})e^{\sqrt{-1}(\alpha,x)}+m(A_{-\alpha}e_{s})e^{\sqrt{-1}(\alpha,x)}\\
 & =\left\{ \left(\sum_{j=1}^{3}\alpha_{j}\sigma_{j}\right)A_{\alpha}+mA_{-\alpha}\right\} e_{s}\cdot e^{\sqrt{-1}(\alpha,x)}\\
 & =\{(|\alpha|d_{\alpha}^{+}+md_{\alpha}^{-})p_{\alpha}+(-|\alpha|d_{\alpha}^{-}+md_{\alpha}^{+})\} e_{s}\cdot e^{\sqrt{-1}(\alpha,x)}\\
 & =\sqrt{E(\alpha)}A_{\alpha}e_{s}\cdot e^{\sqrt{-1}(\alpha,x)}.\end{align*}
Here we use\begin{align*}
|\alpha|d_{\alpha}^{+}+md_{\alpha}^{-} & =\left(|\alpha|+m\sqrt{\frac{\sqrt{E(\alpha)}-|\alpha|}{\sqrt{E(\alpha)}+|\alpha|}}\right)d_{\alpha}^{+}\\
 & =\left(|\alpha|+\frac{m^{2}}{\sqrt{E(\alpha)}+|\alpha|}\right)d_{\alpha}^{+}\\
 & =\sqrt{E(\alpha)}d_{\alpha}^{+}\end{align*}
and\begin{align*}
-|\alpha|d_{\alpha}^{-}+md_{\alpha}^{+} & =\left(-|\alpha|+m\sqrt{\frac{\sqrt{E(\alpha)}+|\alpha|}{\sqrt{E(\alpha)}-|\alpha|}}\right)d_{\alpha}^{-}\\
 & =\left(-|\alpha|+\frac{m^{2}}{\sqrt{E(\alpha)}-|\alpha|}\right)d_{\alpha}^{-}\\
 & =\sqrt{E(\alpha)}d_{\alpha}^{-}.\end{align*}
In addition, we also have\[
m(A_{\alpha}e_{s})e^{\sqrt{-1}(\alpha,x)}+\sqrt{-1}\sum_{j=1}^{3}\sigma_{j}\partial_{j}(A_{-\alpha}e_{s})e^{\sqrt{-1}(\alpha,x)}=\sqrt{E(\alpha)}A_{-\alpha}e_{s}\cdot e^{\sqrt{-1}(\alpha,x)}.\]
This shows that $\phi_{\alpha,s}^{+}$ is an eigenvector for the Dirac
Hamiltonian $\mathfrak{h}$ with an eigenvalue $\sqrt{E(\alpha)}$.
In the same manner, we can prove that $\phi_{\alpha,s}^{-}$ is an
eigenvector for the Dirac Hamiltonian $\mathfrak{h}$ with an eigenvalue
$-\sqrt{E(\alpha)}$.
\end{proof}
Moreover,\[
\{\phi_{\alpha,s}^{\sigma}\ |\ \alpha\in\mathbf{Z}^{3},\  s,\sigma\in\{+,-\}\}\]
 is a C.O.N.S. of $K=\mathbf{C}^{2}\otimes L^{2}(\mathbf{T}^{3})\otimes\mathbf{C}^{2}$. 

Assume $m>1$ (to avoid the complexity) and let $P_{+}$(resp. $P_{-}$)
be a spectral projection of $\mathfrak{h}$ for the positive spectrum
$[m,\infty)$ (resp. negative spectrum $(-\infty,-m]$). Since $\Gamma\phi_{\alpha,s}^{+}=\phi_{-\alpha,s}^{-}$
for $\alpha\in\mathbf{Z}^{3}$ and $s\in\{+,-\}$, $P_{\sigma}$ satisfies
$P_{+}\Gamma=\Gamma P_{-}$. 

Let $H:=P_{+}K$ and $A:=\mathfrak{h}P_{+}$, and let $E$ be the
CH-space constructed from the pair $(H,A)$. 

\begin{prop}
\label{thm:HS_condition_of_massive_n=3D2}$\,$
\begin{enumerate}
\item $\mathrm{ad}(X)$ is not implementable as $\mathcal{L}(\Gamma(H),\mathcal{E}^{*})$,
i.e., $|P_{+}\pi(X)P_{-}\Gamma|_{\mathrm{H.S.}}=+\infty$.
\item $\mathrm{ad}(X)$ is implementable as $\mathcal{L}(\mathcal{E},\mathcal{E}^{*})$.
\end{enumerate}
\end{prop}
We verify the following lemma before proving Proposition \ref{thm:HS_condition_of_massive_n=3D2}.

\begin{lem}
$\,$
\begin{enumerate}
\item Let $\iota:\mathbf{Z}^{3}\to\mathbf{Z}^{3}$ be defined by\[
\iota(\alpha_{1},\alpha_{2},\alpha_{3}):=(-\alpha_{1},\alpha_{2},-\alpha_{3}),\quad\alpha_{i}\in\mathbf{Z}.\]
Then\[
\left(\phi_{\alpha,s}^{+},P_{+}\pi(X)P_{-}\Gamma\phi_{\beta,t}^{+}\right)_{K}=\sqrt{-1}\left(\sigma e_{s},\left(A_{\iota(-\alpha)}A_{\beta}-A_{\iota(\alpha)}A_{-\beta}\right)e_{t}\right)_{\mathbf{C}^{2}}\widehat{\varphi}(\alpha+\beta).\]

\item $A_{\iota(-\alpha)}A_{\beta}-A_{\iota(\alpha)}A_{-\beta}$ equals\[
2(d_{\alpha}^{+}d_{\beta}^{+}-d_{\alpha}^{-}d_{\beta}^{-})(p_{\iota(-\alpha)}p_{\beta}-p_{\iota(\alpha)}p_{-\beta})+2(d_{\alpha}^{+}d_{\beta}^{-}-d_{\alpha}^{-}d_{\beta}^{+})(p_{\iota(-\alpha)}p_{-\beta}-p_{\iota(\alpha)}p_{\beta})\]

\item Put \[
e_{+}:=\frac{1}{\sqrt{2}}\left(\begin{array}{c}
\sqrt{-1}\\
1\end{array}\right),\quad e_{-}:=\frac{1}{\sqrt{2}}\left(\begin{array}{c}
-\sqrt{-1}\\
1\end{array}\right).\]
 (In fact, $\{ e_{\sigma}\}_{\sigma\in\{+,-\}}$ is a C.O.N.S. of
$\mathbf{C}^{2}$.) Then\begin{gather*}
\left(e_{+},(p_{\iota(-\alpha)}p_{\beta}-p_{\iota(\alpha)}p_{-\beta})e_{+}\right)_{\mathbf{C}^{2}}=2\left(\frac{\alpha_{2}}{|\alpha|}-\frac{\beta_{2}}{|\beta|}\right),\\
\left(e_{-},(p_{\iota(-\alpha)}p_{\beta}-p_{\iota(\alpha)}p_{-\beta})e_{+}\right)_{\mathbf{C}^{2}}=2\left\{ \sqrt{-1}\left(\frac{\alpha_{1}}{|\alpha|}+\frac{\beta_{1}}{|\beta|}\right)-\left(\frac{\alpha_{3}}{|\alpha|}+\frac{\beta_{3}}{|\beta|}\right)\right\} \end{gather*}

\item It holds that\begin{gather*}
(d_{\alpha}^{+}d_{\beta}^{+}-d_{\alpha}^{-}d_{\beta}^{-})^{2}=\frac{1}{2}\left(1-\frac{m^{2}}{\sqrt{E(\alpha)E(\beta)}}+\frac{|\alpha||\beta|}{\sqrt{E(\alpha)E(\beta)}}\right),\\
(d_{\alpha}^{+}d_{\beta}^{-}-d_{\alpha}^{-}d_{\beta}^{+})^{2}=\frac{1}{2}\left(1-\frac{m^{2}}{\sqrt{E(\alpha)E(\beta)}}-\frac{|\alpha||\beta|}{\sqrt{E(\alpha)E(\beta)}}\right),\\
(d_{\alpha}^{+}d_{\beta}^{+}-d_{\alpha}^{-}d_{\beta}^{-})(d_{\alpha}^{+}d_{\beta}^{-}-d_{\alpha}^{-}d_{\beta}^{+})=m\left(\frac{1}{\sqrt{E(\beta)}}-\frac{1}{\sqrt{E(\alpha)}}\right).\end{gather*}

\end{enumerate}
\end{lem}
\begin{proof}
(1) can be shown with the help of $\sigma_{2}p_{\alpha}=p_{\iota(\alpha)}\sigma_{2}$.
(2) follows from $d_{\iota(\alpha)}^{\sigma}=d_{\alpha}^{\sigma}$.
(3) is proved by using\begin{gather*}
\sigma_{1}e_{+}=\sqrt{-1}e_{-},\quad\sigma_{2}e_{+}=-e_{+},\quad\sigma_{3}e_{+}=-e_{-}\\
\sigma_{1}e_{-}=\sqrt{-1}e_{-},\quad\sigma_{2}e_{-}=e_{-},\quad\sigma_{3}e_{-}=-e_{+}\end{gather*}
 and\begin{align*}
 & p_{\iota(-\alpha)}p_{\beta}-p_{\iota(\alpha)}p_{-\beta}\\
 & =2\left\{ \left(\frac{\alpha_{1}}{|\alpha|}+\frac{\beta_{1}}{|\beta|}\right)\sigma_{1}-\left(\frac{\alpha_{2}}{|\alpha|}-\frac{\beta_{2}}{|\beta|}\right)\sigma_{2}+\left(\frac{\alpha_{3}}{|\alpha|}+\frac{\beta_{3}}{|\beta|}\right)\sigma_{3}\right\} .\end{align*}
(4) holds by direct computation. 
\end{proof}
\noindent {\it Proof of Proposition} \ref{thm:HS_condition_of_massive_n=3D2}.
Proposition \ref{thm:HS_condition_of_massive_n=3D2} (2) is obvious.
Thus we prove Proposition \ref{thm:HS_condition_of_massive_n=3D2}
(1). We have\begin{align*}
 & \left|\left(\phi_{\alpha,+}^{+},P_{+}\pi(X)P_{-}\Gamma\phi_{\beta,+}^{+}\right)_{K}\right|^{2}+\left|\left(\phi_{\alpha,-}^{+},P_{+}\pi(X)P_{-}\Gamma\phi_{\beta,+}^{+}\right)_{K}\right|^{2}\\
 & =8\ |\widehat{\varphi}(\alpha+\beta)|^{2}\left\{ \left(1-\frac{m^{2}}{\sqrt{E(\alpha)E(\beta)}}\right)+\frac{\alpha_{1}\beta_{1}-\alpha_{2}\beta_{2}+\alpha_{3}\beta_{3}}{\sqrt{E(\alpha)E(\beta)}}\right\} \end{align*}
and hence\begin{align*}
| & P_{+}\pi(X)P_{-}\Gamma|_{\mathrm{H.S.}}^{2}\\
 & =16\sum_{\alpha,\beta\in\mathbf{Z}^{3}}|\widehat{\varphi}(\alpha+\beta)|^{2}\left\{ \left(1-\frac{m^{2}}{\sqrt{E(\alpha)E(\beta)}}\right)+\frac{\alpha_{1}\beta_{1}-\alpha_{2}\beta_{2}+\alpha_{3}\beta_{3}}{\sqrt{E(\alpha)E(\beta)}}\right\} \\
 & =16\ \sum_{\gamma\in\mathbf{Z}^{3}}|\widehat{\varphi}(\gamma)|^{2}\sum_{\alpha\in\mathbf{Z}^{3}}\left\{ 1-\frac{1}{\sqrt{E(\alpha)E(\gamma-\alpha)}}\left(\left(\begin{array}{c}
m\\
\iota(\alpha)\end{array}\right),\left(\begin{array}{c}
m\\
\gamma-\alpha\end{array}\right)\right)_{\mathbf{R}^{4}}\right\} .\end{align*}
(Remark $\iota(\alpha)$, $\gamma-\alpha\in\mathbf{R}^{3}$.) Now
we have the following estimation:\begin{align*}
\sum_{\alpha\in\mathbf{Z}^{3}} & \left\{ 1-\frac{1}{\sqrt{E(\alpha)E(\beta)}}\left(\left(\begin{array}{c}
m\\
\iota(\alpha)\end{array}\right),\left(\begin{array}{c}
m\\
\gamma-\alpha\end{array}\right)\right)_{\mathbf{R}^{4}}\right\} \\
 & \geq\sum_{\alpha\in\mathbf{Z}^{3};\iota(\alpha)=\alpha}\left\{ 1-\frac{\left(\left(\begin{array}{c}
m\\
\alpha\end{array}\right),\left(\begin{array}{c}
m\\
\gamma-\alpha\end{array}\right)\right)_{\mathbf{R}^{4}}}{\left|\left(\begin{array}{c}
m\\
\alpha\end{array}\right)\right|_{\mathbf{R}^{4}}\left|\left(\begin{array}{c}
m\\
\gamma-\alpha\end{array}\right)\right|_{\mathbf{R}^{4}}}\right\} =+\infty.\end{align*}
In fact, if $|\alpha|\to\infty$, then the angle of two vectors $(m,\alpha)\in\mathbf{R}^{4}$
and $(m,\gamma-\alpha)\in\mathbf{R}^{4}$ converges to $\pi$, that
is,\[
\frac{\left(\left(\begin{array}{c}
m\\
\alpha\end{array}\right),\left(\begin{array}{c}
m\\
\gamma-\alpha\end{array}\right)\right)_{\mathbf{R}^{4}}}{\left|\left(\begin{array}{c}
m\\
\alpha\end{array}\right)\right|_{\mathbf{R}^{4}}\left|\left(\begin{array}{c}
m\\
\gamma-\alpha\end{array}\right)\right|_{\mathbf{R}^{4}}}\to-1.\]
Thus $P_{+}\pi(X)P_{-}\Gamma$ is not a Hilbert-Schmidt class operator.
\qed

\subsection{The case of massless ($m=0$) and $n=2$}

Let\begin{gather*}
\phi_{\alpha,s}^{+}(x)=\left(\begin{array}{c}
A_{\alpha}e_{s}\\
A_{-\alpha}e_{s}\end{array}\right)\exp(\sqrt{-1}(\alpha,x)),\\
\phi_{\alpha,s}^{-}(x)=\left(\begin{array}{c}
A_{-\alpha}e_{s}\\
-A_{\alpha}e_{s}\end{array}\right)\exp(\sqrt{-1}(\alpha,x)),\\
\phi_{0,s}^{+}(x):=\frac{1}{\sqrt{2}}\left(\begin{array}{c}
e_{s}\\
e_{s}\end{array}\right),\quad\phi_{0,s}^{-}(x):=\frac{1}{\sqrt{2}}\left(\begin{array}{c}
e_{s}\\
-e_{s}\end{array}\right),\end{gather*}
where $x\in\mathbf{T}^{3}$, $\alpha\in\mathbf{Z}^{3}\backslash\{0\}$,
$s\in\{+,-\}$ and\[
e_{+}:=\frac{1}{\sqrt{2}}\left(\begin{array}{c}
\sqrt{-1}\\
1\end{array}\right),\quad e_{-}:=\frac{1}{\sqrt{2}}\left(\begin{array}{c}
-\sqrt{-1}\\
1\end{array}\right).\]
Moreover, we put \[
A_{\alpha}:=\sqrt{\frac{1}{2}\left(1+\sum_{j=1}^{3}\frac{\alpha_{j}}{|\alpha|}\sigma_{j}\right)}\in\mathrm{Mat}(2,\mathbf{C}),\]
 where $\sigma_{1}$, $\sigma_{2}$, $\sigma_{3}$ are the Pauli matrices.
Then $\phi_{\alpha,s}^{\sigma}$ is an eigenvector for the Dirac operator
$\mathfrak{h}$ with an eigenvalue $\sigma|n|$ and $\{\phi_{\alpha,s}^{\sigma}\ |\ \alpha\in\mathbf{Z}^{3},\  s,\sigma\in\{+,-\}\}$
is a C.O.N.S. of $K=\mathbf{C}^{2}\otimes L^{2}(\mathbf{T}^{3})\otimes\mathbf{C}^{2}$.
In addition, $A_{\alpha}$ is a projection and satisfies $A_{\alpha}A_{-\alpha}=0$. 

Let\begin{gather*}
P_{+}K:=\overline{\mathrm{span}}\{\phi_{n,s}^{+},\ \phi_{0,s}^{+}\ |\  n\in\mathbf{Z}\backslash\{0\},s\in\{+,-\}\},\\
P_{-}K:=\overline{\mathrm{span}}\{\phi_{n,s}^{-},\ \phi_{0,s}^{-}\ |\  n\in\mathbf{Z}\backslash\{0\},s\in\{+,-\}\}\end{gather*}
and $P_{+}$ (resp. $P_{-}$) be a projection from $K$ to $P_{+}K$
(resp. $P_{-}K$). Then $\Gamma\phi_{\alpha,s}^{+}=\phi_{-\alpha,s}^{-}$,
that is, $\Gamma P_{+}=P_{-}\Gamma$. 

Let $H:=P_{+}K$ and $A:=(\mathfrak{h}+cP_{+}-cP_{-})P_{+}$ for some
$c>1$. Then we can define the CH-space $E$ from the pair $(H,A)$. 

\begin{prop}
$\,$
\begin{enumerate}
\item $\mathrm{ad}(X)$ is not implementable as $\mathcal{L}(\Gamma(H),\mathcal{E}^{*})$,
i.e., $|P_{+}\pi(X)P_{-}\Gamma|_{\mathrm{H.S.}}=+\infty$.
\item $\mathrm{ad}(X)$ is implementable as $\mathcal{L}(\mathcal{E},\mathcal{E}^{*})$.
\end{enumerate}
\end{prop}
\begin{proof}
It suffices to show (1). We have\begin{align*}
\sum_{s,t\in\{+,-\}} & \left|\left(\phi_{\alpha,s}^{+},P_{+}\pi(X)P_{-}\Gamma\phi_{\beta,t}^{+}\right)_{K}\right|^{2}\\
 & =16\ |\widehat{\varphi}(\alpha+\beta)|^{2}\left(1+\frac{\alpha_{1}\beta_{1}-\alpha_{2}\beta_{2}+\alpha_{3}\beta_{3}}{|\alpha||\beta|}\right)\\
 & =16\ |\widehat{\varphi}(\alpha+\beta)|^{2}\left\{ 1+\left(\frac{\iota(\alpha)}{|\alpha|},\frac{\beta}{|\beta|}\right)_{\mathbf{R}^{3}}\right\} \end{align*}
for $\alpha$, $\beta\in\mathbf{Z}^{3}\backslash\{0\}$. Thus\begin{align*}
\sum_{\alpha,\beta\in\mathbf{Z}^{3}} & \sum_{s,t\in\{+,-\}}\left|\left(\phi_{\alpha,s}^{+},P_{+}\pi(X)P_{-}\Gamma\phi_{\beta,t}^{+}\right)_{K}\right|^{2}\\
 & \geq16\sum_{\gamma\in\mathbf{Z}^{3}}|\widehat{\varphi}(\gamma)|^{2}\sum_{\alpha\in\mathbf{Z}^{3};\alpha\neq0,\alpha\neq\gamma,\iota(\alpha)=-\alpha}\left\{ 1+\left(\frac{-\alpha}{|\alpha|},\frac{\gamma-\alpha}{|\gamma-\alpha|}\right)_{\mathbf{R}^{3}}\right\} .\\
 & \geq16\sum_{\gamma\in\mathbf{Z}^{3}}|\widehat{\varphi}(\gamma)|^{2}\sum_{\alpha\in\mathbf{Z}^{3};\alpha\neq0,\alpha\neq\gamma,\iota(\alpha)=-\alpha}\left\{ 1+\frac{|\alpha|^{2}-|\alpha||\gamma|}{|\alpha|(|\alpha|+|\gamma|)}\right\} \\
 & \geq16\sum_{\gamma\in\mathbf{Z}^{3}}|\widehat{\varphi}(\gamma)|^{2}\sum_{\alpha\in\mathbf{Z}^{3};|\alpha|>|\gamma|,\iota(\alpha)=-\alpha}\frac{2|\alpha|^{2}}{|\alpha|(|\alpha|+|\alpha|)}\\
 & =+\infty.\end{align*}

\end{proof}

\end{document}